\documentclass{emulateapj}

\usepackage{color}

\usepackage{graphicx}

\newcommand{\chan}{{\em Chandra}}
\def\psr {PSR\, B0540$-$69}

\shorttitle{UV Observations of PSR\, B0540$-$69}
\shortauthors{Mignani et al.}

\begin{document}

\title{The first ultraviolet detection of the Large Magellanic Cloud pulsar PSR\, B0540$-$69 and its multi-wavelength properties}

\author{
R. P. Mignani\altaffilmark{1,2}, 
A. Shearer\altaffilmark{3},
A. de Luca\altaffilmark{1},
F. E. Marshall \altaffilmark{4},
L. Guillemot \altaffilmark{5},
D. A. Smith \altaffilmark{6},
B. Rudak\altaffilmark{7}, 
L. Zampieri\altaffilmark{8},
C. Barbieri\altaffilmark{8,9,10},
G. Naletto\altaffilmark{9},
C. Gouiffes\altaffilmark{11},
G. Kanbach\altaffilmark{12}
}

\affil{\altaffilmark{1} INAF - Istituto di Astrofisica Spaziale e Fisica Cosmica Milano, via E. Bassini 15, 20133, Milano, Italy}
\affil{\altaffilmark{2} Janusz Gil Institute of Astronomy, University of Zielona G\'ora, ul Szafrana 2, 65-265, Zielona G\'ora, Poland}
\affil{\altaffilmark{3} Centre for Astronomy, School of Physics, National University of Ireland Galway, University Road, Galway, Ireland}
\affil{\altaffilmark{4} Astrophysics Science Division, NASA Goddard Space Flight Center, Greenbelt, MD 20771, USA}
\affil{\altaffilmark{5} Laboratoire de Physique et Chimie de l'Environnement et de l'Espace (LPC2E), CNRS-Universit\'e d'Orl\'eans, F-45071 Orl\'eans, France}
\affil{\altaffilmark{6} Centre d'Etudes Nucléaires de Bordeaux Gradignan, IN2P3/CNRS, Universit\'e de Bordeaux 1, BP120, F-33175, Gradignan Cedex, France}
\affil{\altaffilmark{7} Nicolaus Copernicus Astronomical Center, Polish Academy of Sciences, ul. Rabia\'nska 8, 87100, Toru\'n, Poland }
\affil{\altaffilmark{8}INAF - Osservatorio Astronomico di Padova, Vicolo dell'Osservatorio 5, I-35122 Padova, Italy}
\affil{\altaffilmark{9} Department of Physics and Astronomy "G. Galilei", University of Padova, Vic. Osservatorio 3, I-35122 Padova, Italy}
\affil{\altaffilmark{10} Centro di Ateneo di Studi ed Attivit\'a Spaziali "Giuseppe Colombo", University of Padova, Via Venezia 15, I-35131 Padova, Italy}
\affil{\altaffilmark{11} Laboratoire AIM, UMR 7158 (CEA/Irfu, CNRS/INSU, Universit\'e Paris VII), CEA Saclay, Bt. 709,F-91191 Gif-sur-Yvette Cedex, France}
\affil{\altaffilmark{12} Max-Planck Institut f\"ur Extraterrestrische Physik, Giessenbachstrasse 1, 85741, Garching bei M\"unchen, Germany}

\begin{abstract}
We observed the young ($\sim 1700$ yrs) pulsar \psr\ in the near-ultraviolet (UV) for the first time with the Space Telescope Imaging Spectrograph (STIS) aboard the {\em Hubble Space Telescope}.  Imaging observations with the NUV- and FUV-MAMA detectors in TIME-TAG mode allowed us to clearly detect the pulsar in two bands around 2350\AA\ and 1590\AA, with magnitudes $m_{\rm NUV} =21.45 \pm 0.02$ and $m_{\rm FUV} =21.83 \pm 0.10$. 
We also detected the pulsar-wind nebula (PWN) in the NUV-MAMA image,  with a morphology  similar to that observed in the optical and near-infrared (IR). 
The extinction-corrected NUV and FUV pulsar fluxes are compatible with a very steep power law spectrum $F_{\nu} \propto \nu^{-\alpha}$ with spectral index $\alpha_{\rm UV} \sim 3$,  non compatible with a Rayleigh Jeans spectrum, indicating a non-thermal origin of the emission. The comparison with the optical/near-IR power-law spectrum (spectral index $\alpha_{\rm O,nIR} \sim 0.7$), indicates an abrupt turnover at wavelengths below 2500 \AA,  not yet observed in other pulsars.
We detected pulsations in both the NUV and FUV data at the 50 ms pulsar period. 
In both cases, the folded light curve features a broad pulse with two peaks closely spaced in phase, as observed in the optical and X-ray light curves. The NUV/FUV peaks are also aligned in phase with those observed in the radio (1.4 GHz), optical, X, and $\gamma$-ray light curves, like in the Crab pulsar,  implying a similar beaming geometry across all wavelengths.   
\psr\ is now  the fifth isolated pulsar, together with Crab, Vela, PSR\, B0656+14, and the radio-quiet Geminga, detected in the optical, near-UV, near-IR, X-rays and $\gamma$-rays, and seen to pulsate in at least four of these energy bands.
\end{abstract}


\section{Introduction}

Pulsars are rapidly spinning neutron stars that emit electromagnetic radiation (mostly) at the expenses of their rotational energy (Pacini 1968; Gold 1968), hence also referred to as rotation-powered pulsars. Apart from the radio band, where the first of the over 2500 radio pulsars known to date\footnote{See, ATNF pulsar catalogue (Manchester et al.\ 2005)} was originally discovered (Hewish et al.\ 1968), pulsars 
are also observed
in X-rays, $\gamma$-rays, optical, infrared (IR), ultraviolet (UV), and the sub-mm 
(Mignani et al.\ 2017).

Owing to their intrinsic faintness, the number of pulsar detections at optical energies by and large lag behind those at 
high energies.
After the Crab pulsar (PSR\, B0531+21; Cocke et al.\ 1969), which was the first one identified through its optical pulsations at the radio period (Cocke et al.\ 1969), only  eight isolated pulsars
 (i.e., not in binary systems) 
have been firmly identified in the optical plus two candidates (see, Mignani 2011 for a review), and three more identifications have been recently proposed (Moran et al.\ 2013; Mignani et al.\ 2016a; Rangelov et al.\ 2017). Optical pulsations have been detected only for some of them, though. Indeed,  apart from the Crab (Cocke et al.\ 1969), only for four other pulsars have optical pulsations been detected: the Vela pulsar (PSR\, B0833$-$45; Wallace et al.\ 1977),  PSR\, B0540$-$69 (Middleditch \& Pennypacker 1985),  PSR\, B0656+14 (Shearer et al.\ 1997) and Geminga (Shearer et al.\ 1998).  Eight of the 
isolated pulsars 
identified in the optical have also been detected in the UV 
with the {\em Hubble Space Telescope} ({\em HST})
and four of them (Crab, Vela, PSR\, B0656+14, and Geminga)
pulsate in the UV (Percival et al.\ 1993; Romani et al.\ 2005; Shibanov et al.\ 2005; Kargaltsev et al.\ 2005), beside the optical band.
These four pulsars have also been identified in the near-IR
(Mignani et al.\ 2012 and references therein), but 
pulsations in this band have been detected only for the Crab  (e.g., Eikenberry et al.\ 1997).

The near-UV/optical/near-IR (hereafter UVOIR) spectra of young pulsars ($\tau_{\rm C} \la 10$ kyrs), where $\tau_{\rm C}$ is the characteristic age\footnote{This is defined as $P_{\rm s}/ (2 \dot{P_{\rm s}})$, where $P_{\rm s}$ and $\dot{P_{\rm s}}$ are the pulsar spin period and its first derivative, respectively.}, show the signature of non-thermal, likely synchrotron, emission from the neutron star magnetosphere (see, e.g. Mignani 2011), characterised by power-law (PL) spectra $F_{\nu} \propto \nu^{-\alpha}$ ($\alpha \sim$0--1). Differences in the spectral index $\alpha$ 
across the three bands 
are observed in some cases, e.g. the Crab pulsar (Sollerman 2003), but not in others, e.g. the Vela pulsar (Zyuzin et al.\ 2013). In middle-aged pulsars ($\tau_{\rm C}  \approx 0.1$--1  Myr), a second emission component is present in the optical/UV, associated with thermal emission from the neutron star surface and characterised by a Rayleigh-Jeans (RJ)  spectrum
with a brightness temperature $T_{\rm B} \approx10^{5}$ K (Mignani 2011).

Despite the optical and UV being very close in wavelengths, differences in the pulsar light curves\footnote{Through the text we implicitly refer to the light curves folded at the pulsar spin period.}  exist. In particular, {\em HST} observations showed that the widths and separations of the two peaks in the Crab light curve  (a.k.a. Main Pulse and Interpulse) are larger in the optical than in the UV (Percival et al.\ 1993), perhaps related to the difference in the PL slope  between these two bands (Sollerman 2003).  In the case of Vela  possible differences in the widths and separations of the two main peaks between the optical and the UV light curves cannot be appreciated owing to the lower statistics,  although they differ in the structure of the smaller peaks (Romani et al.\ 2005). At variance with the Crab, there is no difference in the PL slope from the optical to the UV (Zyuzin et al.\ 2013).  {\em HST} observations of the middle-aged pulsars PSR\, B0656+14 (Shibanov et al.\ 2005) and Geminga (Kargaltsev et al.\ 2005) also showed differences in their light curves from the optical to the UV. This might also be due to the rising contribution of the RJ component in the UV relative to the PL component, with extra modulations possibly produced by hot spots on the neutron star surface.
Detecting more optical/UV pulsars is important to study the evolution of the light curve and spectrum across these two bands and to infer the characteristics and geometry of the corresponding emission regions.

\psr\  in the Large Magellanic Cloud (LMC) is  the second brightest optical pulsar  ($V=22.5$) after the Crab and
an obvious target for UV observations,  which have thus far not been performed.  It is  referred to  as the  Crab ``twin''  because it  is  very  similar in  spin period  ($P_{\rm s}= 50$  ms), period derivative ($\dot{P_{\rm s}} \sim 4.78 \times 10^{-13}$s s$^{-1}$), characteristic age  ($\tau_{\rm C} \sim 1.7$  kyr), rotational energy loss ($\dot {E}  \sim 1.5 \times 10^{38}$ erg~s$^{-1}$), and surface magnetic field ($B_{\rm s} \sim 4.98 \times 10^{12}$ G)\footnote{The latter two values have been derived from the standard  formulae $\dot{E} = 4 \times 10^{46}  \dot{P}_{\rm s}/P_{\rm s}^{3}$ erg s$^{-1}$ and $B = 3.2 \times 10^{19} \sqrt{P_{\rm s} \dot{P}_{\rm s}}$ G, derived by assuming for the neutron star a moment of inertia $I = 10^{45}$ g cm$^{2}$.}. 

The LMC distance (48.97$\pm$0.09 kpc; Storm et al.\ 2011) makes \psr\ one of the faintest radio pulsars 
(Manchester et al.\ 1993).
Indeed,  it  was
discovered in X-rays 
(Seward et al.\ 1984) becoming the first extragalactic pulsar detected at any wavelength.  
\psr\ is the latest pulsar to have been detected in the near-IR (Mignani et al.\ 2012)
and has also been 
recently detected as a $\gamma$-ray pulsar  (Ackermann et al.\ 2015) by the {\em Fermi} Large Area Telescope (LAT).
Like other young pulsars (Kargaltsev et al.\ 2017), it is embedded in a bright pulsar wind nebula (PWN) visible from the near-IR  to the soft/hard X-rays (Mignani et al.\ 2012; Petre et al.\ 2007; S\l{}owikowska et al.\ 2007).
Optical pulsations 
were 
detected by Middleditch \& Pennypacker (1985),
while the pulsar counterpart was later identified via high-resolution imaging (Caraveo et al.\ 1992; Shearer et al.\ 1994).
The
optical light curve  (Middleditch et al.\ 1987;  Gouiffes et al.\ 1992; Boyd et al.\  1995) features a broad pulse, which is actually
resolved in two  peaks
(see also, Gradari et al.\ 2011). 

The optical 
spectrum of \psr\ 
is characterised by 
a PL   (e.g., Serafimovich et al.\ 2004).
The measurement of significant phase-averaged polarisation with the {\em HST} (Mignani et al.\ 2010a; Lundqvist et al.\ 2011) confirmed the magnetospheric origin of its optical emission. 
{\em HST}  and Very Large Telescope (VLT) adaptive optics images (Mignani et al.\ 2010a; 2012) clearly resolved \psr\ from its  compact (4\arcsec) PWN, making it possible to precisely measure the pulsar flux.
This 
yielded the most accurate measurement of its PL spectral index in the optical/near-IR ($\alpha_{\rm O,nIR}= 0.70 \pm 0.04$), which is similar to that in the X-rays ($\alpha_{\rm X}=0.83\pm 0.13$), measured from  \chan\ spectroscopy (Kaaret et al.\ 2001). However, the optical fluxes fall below the extrapolation of the X-ray PL (Mignani et al.\ 2010a), suggesting a spectral 
flattening
in the UV. 
Determining the pulsar spectrum in the UV is, then, key to 
confirm the expected flattening,
whereas measuring the UV light curve is key to determine whether such flattening is associated with different optical and UV light curve profiles, as possibly observed in the Crab pulsar (Percival et al.\ 1993).

Here, we present the results of the first UV observations of \psr, carried out with the {\em HST}. 
This manuscript is organised as follows: observations and data analysis are described in Section 2, whereas the results are presented and discussed in Sections 3 and 4, respectively. Summary and conclusions follow in Section 5.

\section{Observations}

\subsection{Observation Description}

We observed \psr\ with the  {\em HST} during Cycle 23 (Prog. ID: 14250; PI: Mignani) on February 27 and 28 2017,  as part of  the  UV Initiative Program. 
We used the Space Telescope Imaging Spectrograph (STIS) and collected data with both its NUV- and FUV-MAMA (Multi-Anode Micro-channel Array) detectors that are sensitive in the 1600--3100\AA\ and 1150--1700\AA\ spectral ranges, respectively. 

The detectors were operated in imaging ($25\arcsec \times 25\arcsec$ field--of--view) TIME-TAG mode, chosen  for two principal reasons: (i) to clearly resolve the pulsar emission from that of the surrounding  PWN ($\sim 4\arcsec$ diameter), thanks to a spatial resolution of 0\farcs024/pixel, and  (ii) to search for pulsations at the pulsar period  (50 ms) and  accurately sample the light curve, thanks to a time resolution of $125 \mu$s. For the NUV- and FUV-MAMA detectors we used their F25QTZ 
filters\footnote{Since the filters have the same name for both MAMA detectors, hereafter we simply distinguish the two data sets by the detector name (NUV and FUV for short).}, which have central wavelengths and FWHM bandwidths of $\lambda=2359.3$\AA, $\Delta \lambda=998.7$\AA\ and $\lambda=1596.2$\AA, $\Delta \lambda=231.6$\AA,  respectively (Riley et al.\ 2017). These combinations provide high-throughput broad-band UV imaging, minimising the background contribution from geo-coronal emission lines and maximising the  spectral coverage achievable with the STIS MAMAs.  

The planned exposures were allocated in six spacecraft orbits, equally distributed between the NUV and FUV observations (three orbits each), and split in two different visits to cope with the {\em HST} scheduling constraints.  The same roll angle of 32\fdg4934  (measured east of north) was used in both visits.    The exposure time per orbit was defined to fully exploit the target visibility window\footnote{Owing to scheduling constraints in Cycle 23 it was not possible to observe our target in Continuous Viewing Zone and the maximum visibility window before Earth occultation was about 3500 s per orbit.}. After accounting for instrument overheads and guide star acquisitions, we acquired one 3050 s and two 3300 s exposures (1000 s buffer time)  in each visit for a net total integration time of 9650 s for both the NUV and FUV observations.  

\subsection{Data Analysis}

We retrieved our data from the Mikulski Archive for Space Telescopes (MAST\footnote{{\tt https://archive.stsci.edu/hst/search.php}}) after routine data reduction and calibrations steps  have been applied through the {\tt CALSTIS} pipeline under the {\sc STSDAS} package. These steps are: dark subtraction, flat fielding, geometric distortion and detector non-linearity corrections, flux calibration,  which have all been implemented using the calibration frames and tables closest-in-time to our observations.  In order to increase the signal--to--noise, for each data set we then co-added the single exposures using the {\sc STSDAS} task {\tt combine} which also applies rejection of  cosmic ray hits. 

We checked the astrometry of the NUV and FUV images, determined by the {\em HST} aspect solution, against the {\em HST}/WFPC2 images of Mignani et al.\ (2010a), whose astrometry was re-calibrated in the 2MASS (Skrutskie et al.\  2006) reference frame with an overall accuracy of 0\farcs12--0\farcs15.   To account for the measured absolute offsets of $\sim 0\farcs23$ and $\sim 0\farcs57$ in the NUV and FUV image  astrometry,  respectively, we used a grid of reference stars detected in a  $\sim 12\arcsec$ radius  around \psr\   to register the MAMA images onto the astrometry reference frame of the WFPC2 images with an accuracy of better than 0\farcs01. 
We used the 
NUV and FUV images with the re-calibrated astrometry as a reference for the pulsar identification.

\section{Results}

\subsection{Imaging and Photometry}

Fig.\ 1 shows the NUV- and FUV-MAMA images of the \psr\ field obtained after the processing described in the previous section. We clearly detected \psr\ in both the NUV and FUV images at a position coincident
with its optical coordinates computed by Mignani et al.\ (2010a): $\alpha_{J2000}=05^h  40^m  11\fs202$ (0\fs009), $\delta_{J2000}=  -69^\circ   19\arcmin  54\farcs17$ (0\farcs05)\footnote{The errors refer to the average of the pulsar coordinates computed on four independent WFPC2 data sets, see Mignani et al.\ (2010a) for details.}. Ours is the first detection ever of \psr\ in the UV, which also makes it the fifth  isolated pulsar, among the $\sim 2300$ known, that has been detected in the near-IR, optical, UV, X-rays and $\gamma$-rays, after the Crab and Vela pulsars, PSR\, B0656+14, and Geminga (the fourth among radio pulsars\footnote{Geminga (PSR\, J0633+1746) has not yet been unambiguously detected as a radio pulsar despite many searches, see Maan (2015) and references therein.}).
 We also detected the \psr\ PWN  in our NUV image (Fig.\ 1, left), with a structure and extent similar to what is observed in the optical and near-IR (Mignani et al.\ 2010a; 2012). The PWN  is at most barely visible, however, in the FUV image (Fig.\ 1, right). The bright emission knot in the PWN detected at $\sim 1\farcs7$  south west of the pulsar  in {\em HST}/WFPC2 images (De Luca et al.\ 2007) is also visible in the STIS/NUV-MAMA observation, aligned with the major axis of the PWN. 
  Since our paper is focused on the pulsar, a coherent multi-wavelength spectral and spatial analysis of the PWN and its features
  will be the subject of a subsequent paper.

\begin{figure*}
\centering
{\includegraphics[width=5.5cm,angle=90,clip=]{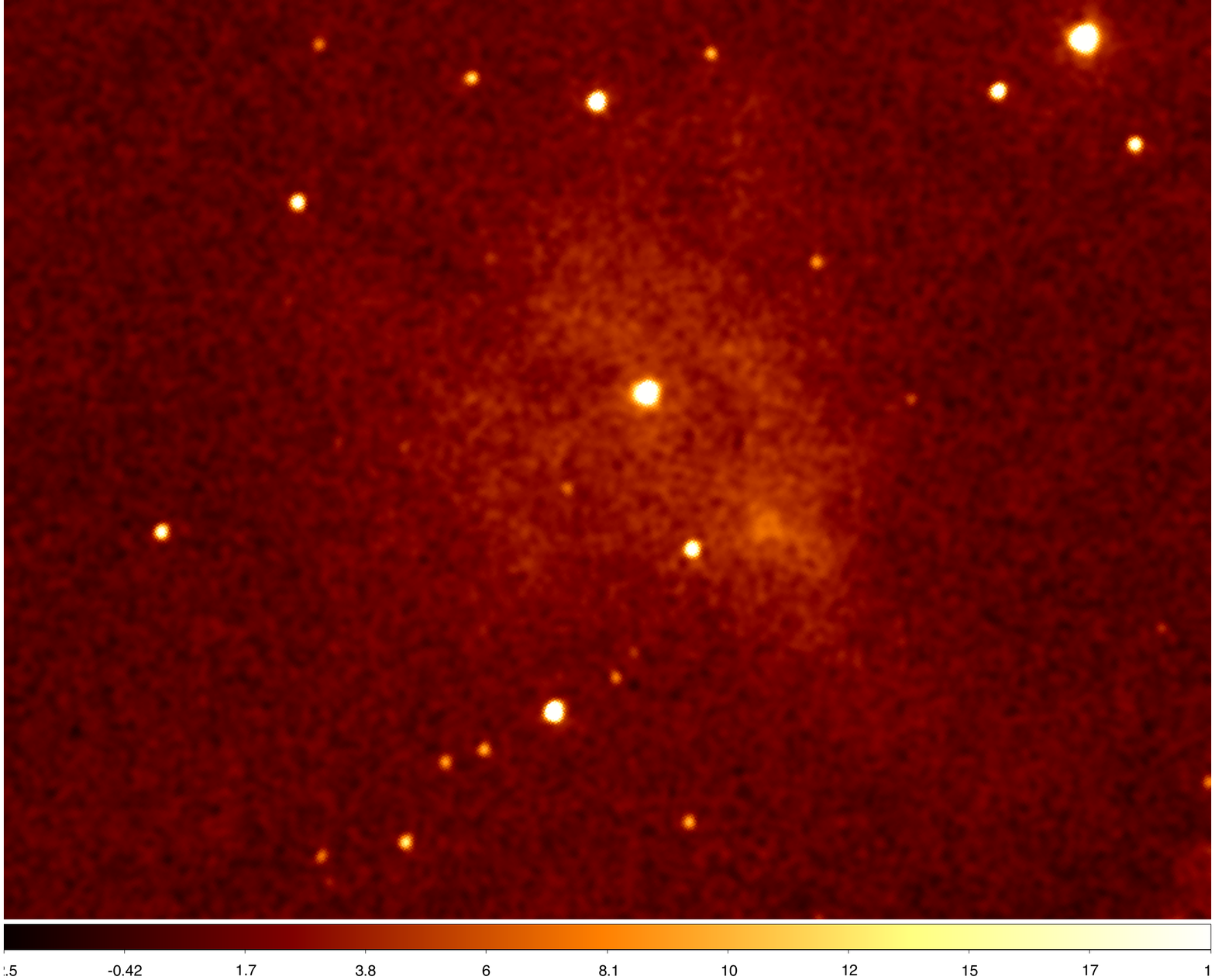}}
{\includegraphics[width=5.5cm,angle=90,clip=]{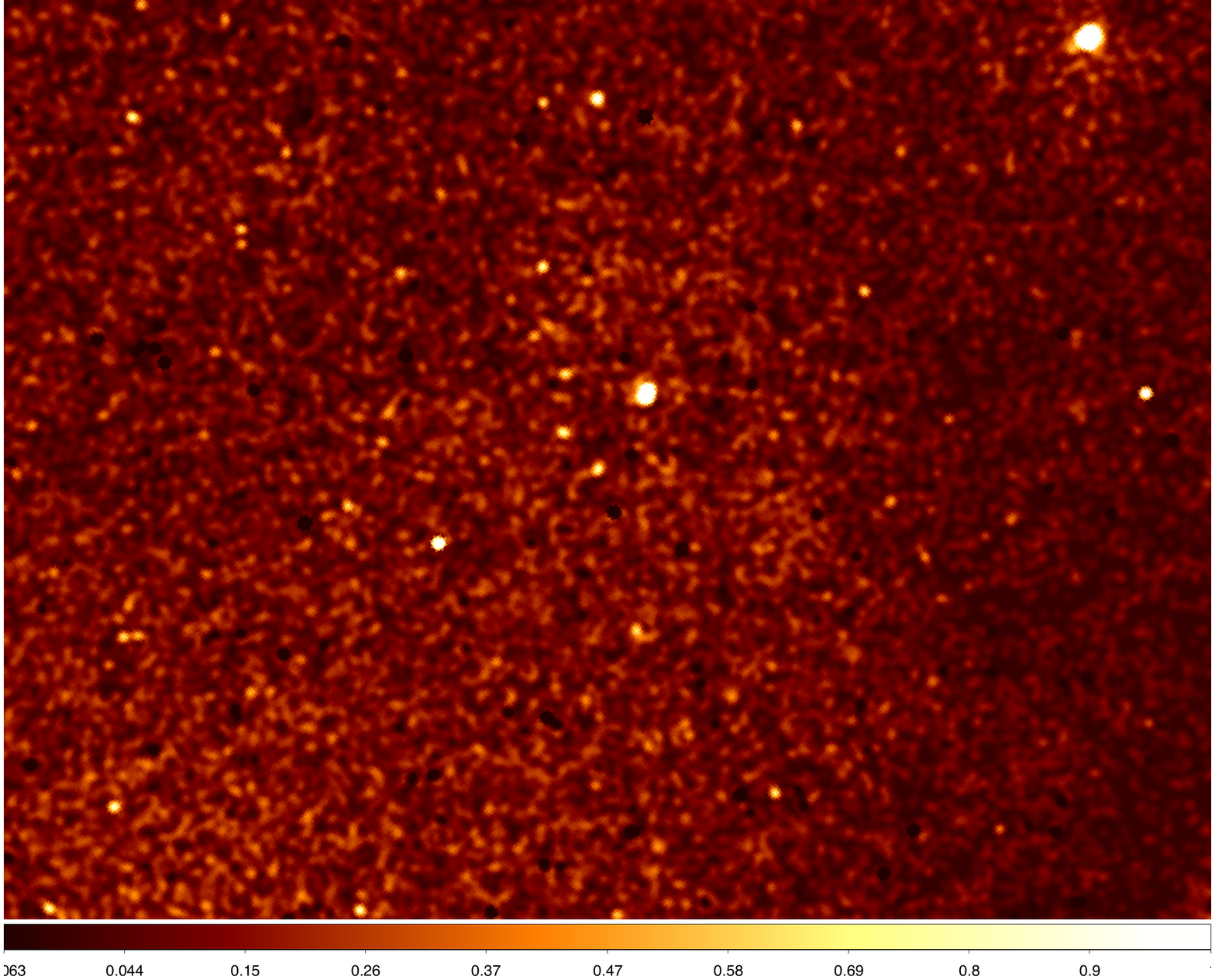}}
\caption{\label{ima} STIS NUV- (left) and FUV- (right) MAMA images (9650 s each) of the \psr\ field taken with the F25QTZ filters. The images are aligned in right ascension and declination, with North to the top and east to the left. The image size is 9\arcsec$\times$9\arcsec. The pulsar is the source at the centre of the nebula, clearly visible in the NUV-MAMA image.  The intensity scale of the two images (in counts) is colour-coded in the two horizontal bars at the bottom of each image. In both cases, the minimum/maximum of the scale  have been adjusted to favour the identification of the pulsar and other stars in the field.   For a better visualisation, both images have been smoothed with a Gaussian function using a kernel of 2 pixel radius.
}
\end{figure*}

For both the NUV and FUV images, we computed the pulsar fluxes through aperture photometry employing the tools in the {\sc IRAF}\footnote{IRAF is distributed by the National Optical 
Astronomy Observatories, which are operated by the Association of Universities for Research in Astronomy, Inc., under cooperative agreement with the National Science Foundation.}  package {\em PHOT}.  We used an aperture radius of 10 pixel 
(0\farcs24) to maximise the signal--to--noise and we sampled the sky background within an annulus of 25 pixel inner radius (0\farcs6) to avoid contamination from the wings of the pulsar PSF, which are particularly bright in the NUV-MAMA F25QTZ filter, and of 35 pixel outer radius (0\farcs84) to avoid including bright stars close to the pulsar. Since the PWN is the main source of background, which is itself not spatially uniform, we carefully checked that our photometry is not very sensitive to the choice of the annulus width.  We then applied the aperture correction to compute the pulsar count-rates in an infinite aperture using the values of the encircled energy fractions for the chosen radius reported in the  STIS Instrument Handbook Version 16.0 (Riley et al.\ 2017) for the NUV- and FUV-MAMA  F25QTZ filters. The aperture-corrected, background-subtracted count-rates (CR) are 1.53$\pm$0.02 
and 0.053$\pm$0.005 counts s$^{-1}$ for the NUV and FUV images, respectively. The large difference in CR can be visually appreciated by the comparison of the two images (Fig. \ref{ima}), which were obtained through the same integration time (9650 s) and are, thus, directly comparable to each other. We converted the corresponding instrumental magnitudes into ST magnitudes (STMag) using the photometric calibration parameter {\sc PHOTFLAM}, which is closest in time  to our observations and reported in the image header, according to the definition: STMag$ = -2.5 \times log_{10} ({\rm CR} \times {\sc PHOTFLAM}) - 21.10$. The observed magnitudes, i.e. uncorrected for the interstellar extinction,  are $m_{\rm NUV} =21.45 \pm 0.02$ and $m_{\rm FUV} =21.83 \pm 0.10$, where the associated errors are purely statistical.

\subsection{Timing}

We looked for pulsations in the NUV and FUV data of \psr\ at its $\sim$50 ms period. We extracted the time series from the event files using an aperture with a radius of 10 pixels (0\farcs24), which corresponds to 82\% of the pulsar flux (Riley et al.\ 2017).
Then, we used the task {\tt hstephem} in {\sc STSDAS}  to account for the spacecraft position and velocity during the observations, and the {\sc IRAF} task {\tt otimedelay} to convert the photon arrival times from the topocentric reference frame to the solar system barycentre.  As a reference for the pulsar position we used the most precise coordinates known (Mignani et al.\  2010a). Since the pulsar has a proper motion of $< 1$ mas yr$^{-1}$ (Mignani et al.\ 2010a) any displacement between the epoch of our STIS observations (MJD 57811) and that of the reference position (MJD 54272) is much smaller than the absolute uncertainty on the pulsar coordinates (70 mas).

We folded the NUV and FUV time series around the expected pulsar period using as a reference the most recent timing solution for \psr\ (Marshall et al., in preparation), obtained by monitoring the pulsar period evolution with the X-ray Telescope (XRT; Burrows et al.\ 2005) aboard the {\em Neil Gehrels Swift Observatory} after the change in the spin-down frequency derivative $\dot{\nu}$ occurred between December 3 2011 and December 17 2011 (Marshall et al.\ 2015).  Results from this monitoring program, which started on February 17 2015, have been presented in Marshall et al.\ (2016) and Marshall et al.\ (2018).  
In particular, the data set consists of all the XRT observations of \psr\ that covers the full time span from February 17 2015 to March 28 2018 (MJD 57070--58205), for a total of 176.893 ks. 

The analysis of the data from the XRT  used the same procedure described in Marshall et al.\ (2015) and Marshall et al.\ (2016).  All the observations were  made using the Window Timing mode. The data were processed using the software tool {\tt xrtpipeline}, and events were screened to maximize the signal from the pulsar. Arrival times were corrected to the solar system barycentre using the pulsar position from Mignani et al.\ (2010a) and the JPL Planetary Ephemeris DE-200 (Standish et al.\ 1982)\footnote{In this, we followed the prescription of the {\sc HEASoft} tool
{\tt barycorr} that recommends DE-200 for the {\em Swift} data, which are based on the FK5 reference frame. More recent ephemeris files, such as DE-405, use the ICRS reference frame. According to the notes to the {\tt barycorr} tool, using DE-405 instead of DE-200 will cause a maximum error of 2 ns for satellites in low-Earth orbit, which is completely negligible for \psr. Moreover the STIS data analysis threads recommend the use of DE-200.}. Events were folded on multiple candidate periods, and a sine wave was fit to the best folded light curve. The resulting frequencies and phases at the epochs of the observations were then fit with a spin model  using a Taylor expansion of spin frequency and its derivatives through $\ddot{\nu}$. Small glitches in the pulsar on MJD 57546 and 57946  were added to the model to produce a good fit for the entire span of observations. The resulting ephemeris was used to produce a folded X-ray light curve of all the events. The updated timing solution, together with a detailed analysis of the spin frequency evolution, will be presented in a separate publication (Marshall et al., in preparation).

\begin{figure*}
\centering
{\includegraphics[width=12cm]{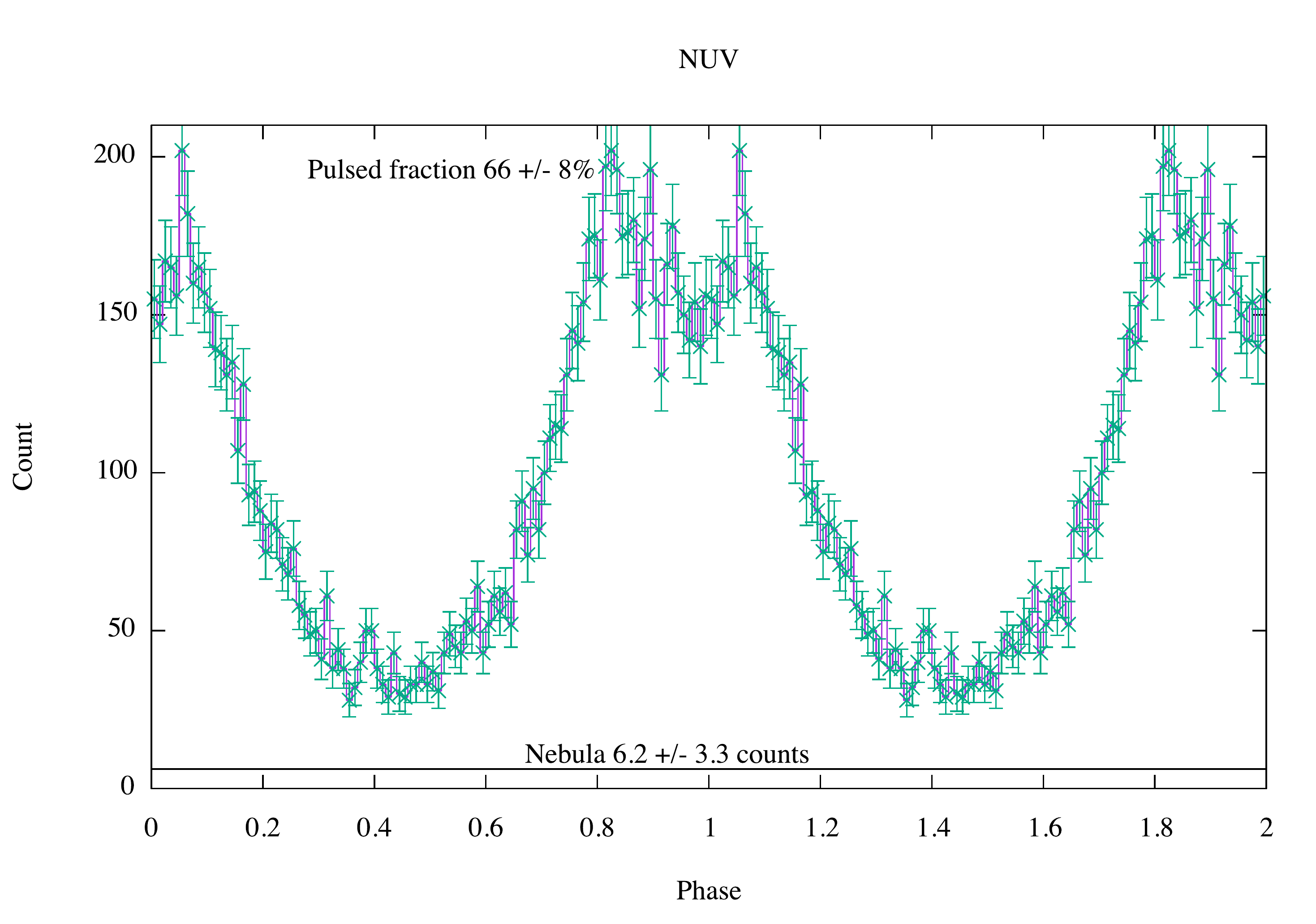}}
{\includegraphics[width=12cm]{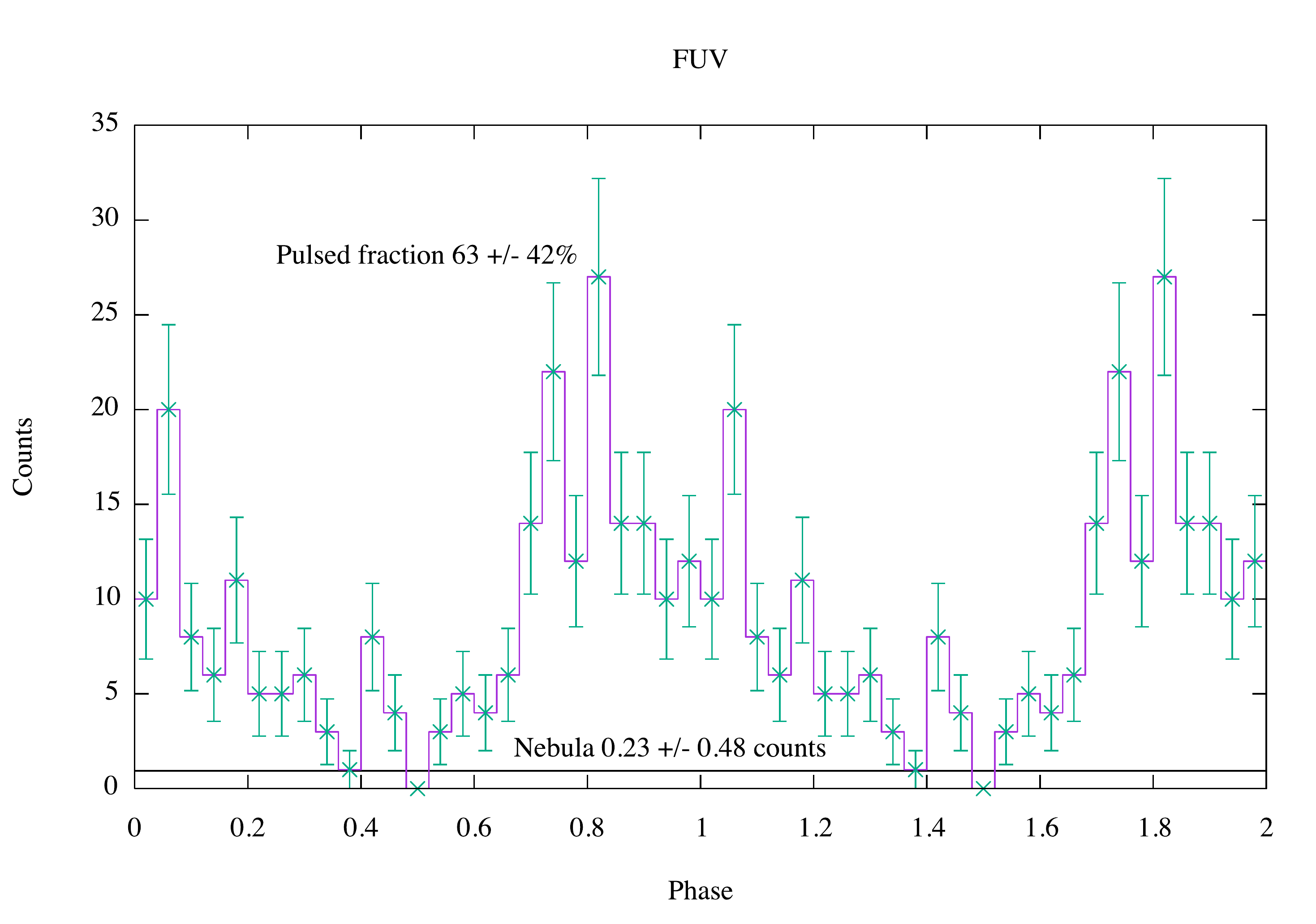}}
\caption{\label{lc} NUV (top) and FUV (bottom) light curves of \psr. Two cycles are shown for clarity. Owing to the difference in signal--to--noise, a different phase binning has been applied in each case.   
The NUV and FUV light curves have been folded using the most recent ephemeris obtained from {\em Swift}/XRT observations over the period February 17 2015--March 28 2018 (Marshall et al., in preparation). Pulsar counts have been extracted using an aperture of 10 pixel radius (0\farcs24).  In both panels, the horizontal dashed line marks the background level of the PWN computed in annulus centred on the pulsar of 25 pixel inner radius and 10 pixel width (\S\ 4.2).
}
\end{figure*}

By folding the NUV and FUV data at the expected pulsar period, we found a clear pulsed signal  in both the NUV and FUV time series (Fig. \ref{lc}a,b), albeit at a different significance level owing to the difference in the pulsar count-rate in the two data sets (see \S 3.1). The detection significance for the NUV and FUV pulsations is
$\sim 36 \sigma$ and $\sim6 \sigma$
respectively,  which we computed based on the 
$Z^2_n$  statistic (Buccheri et al.\ 1983). We also computed the detection significance based on a 
$\chi^2$ analysis and obtained comparable results, with $\chi^2$ values of 2800 (64 d.o.f.) and 82 (20 d.o.f.) for the NUV and FUV light curves,  respectively. The final probability was determined with Monte Carlo simulations for the case of non-normal distributions. The detection of the expected periodicity clearly and independently confirms the pulsar UV identification, initially based upon position match with the optical coordinates. Therefore, \psr\ is also the fifth isolated pulsar
for which pulsations have been detected in the UV, optical, X and $\gamma$-rays. A comprehensive cross-comparison of the multi-wavelength light curves and spectra of these five pulsars is beyond the goals of this work and will be reported elsewhere.

Both the folded 	NUV and FUV light curves feature a broad pulse resolved in two peaks separated by $\sim$0.3 in phase (Fig.\ 2).   The two peaks in the NUV light curve look more structured than in the FUV one, but this is only an effect of  the better count statistics and smaller binning.  
Although similar, the NUV and FUV light curves show some small differences.  For instance, in the FUV light curve the intensity of the second peak seems to be lower than the first one, whereas in the NUV one the intensity of the two peaks is comparable.   Obtaining a FUV light curve with an improved signal--to--noise would help to determine whether this difference in the intensity of the two peaks is real or it is an effect of the low count statistics. A difference between the  relative intensity of the two main peaks is also observed, e.g. in the STIS NUV and FUV light curves of the Vela pulsar, where the intensity of the primary peak with respect to the secondary one increases in the FUV (Romani et al.\ 2005). On the other hand, in the case of the Crab pulsar there is no appreciable difference in the relative intensity of the primary and secondary peaks between the STIS NUV and FUV light curves (Sollerman et al.\ 2000). A hint of a third peak between the two main ones is visible in the NUV light curve of \psr, which looks more or less prominent depending on the binning. Its estimated significance is $\sim$ 2--3 $\sigma$ only and it is even lower in the FUV light curve, where such a third peak is barely visible. A possible precursor to the broad pulse is also visible in the FUV light curve, but not as clearly in the NUV one, indicating that it might be an effect of the different binning.  Also in this case, obtaining NUV  and FUV light curves with  an improved signal--to--noise  would help to assess the existence of these features. This complex structure might be reminiscent of that observed in the NUV and FUV light curves of the Vela pulsar, where two small peaks are present in addition to the primary and secondary ones (Romani et al.\ 2005). At present,  however, we cannot determine whether the sub-structures seen in the \psr\ light curves are real or not. Investigations through deeper observations may be worthwhile.

As it can be seen from Fig.\ref{lc},  the signal from \psr\  is almost totally pulsed in the NUV and FUV light curves.
In particular, the  pulsed fraction (PF)  is 66\%$\pm$8\% and 63\%$\pm$42\% in the NUV  and FUV light curves, respectively, where we computed PF from the component above the flux level in the phase interval defined as the off-pulse region, often refereed to as the direct current (DC) level.  The DC level is much higher in the NUV light curve than in the FUV one (Fig. \ref{lc}), as expected from the higher PWN background present in the NUV image (Fig. \ref{ima}). This, however, only partially accounts for the DC level, which is well above the nebula background.  Indeed,  the DC level, calculated in the  phase intervals 0.355--0.515 (NUV) and 0.34--0.54 (FUV), is 35.8$\pm$6.9 counts and 3.2$\pm$2.8 counts, respectively.  On the other hand, the nebula background, which we sampled in annulus centred on the pulsar of 25 pixel inner radius and 10 pixel width (See \S\ 3.1),  is only 6.2$\pm$3.3 counts and 0.23$\pm$0.48 counts in the NUV and FUV images, respectively. 

Such a difference suggests that there is a significant continuous emission component from a source very close to the pulsar, which is not associated with the PWN.
A significant DC component is also seen in the Crab pulsar light curve and is associated with the unresolved emission from the bright knot in the PWN, at 0\farcs65 from the pulsar (S\l{}owikowska et al.\ 2009). No such structure, however, is seen in our high-spatial resolution {\em HST} images of \psr\ (see, also Mignani et al.\ 2010). Given the small aperture used to extract the pulsar counts (0\farcs24 radius), about twice the size of the image PSF,  we deem it unlikely that this DC component is associated with a source other than the pulsar itself. Therefore, there must be an emission component from the pulsar which is not pulsed. This would be the case, for instance, if such a component were emitted isotropically from the pulsar magnetosphere, within the last closed magnetic field lines. Time-resolved UV spectroscopy observations are needed to verify this hypothesis by studying the pulsar spectrum as a function of the  rotation phase.  In this way, it would be possible to determine whether the spectrum of the DC component differs from that of the pulsed component and, more importantly, whether it varies with the rotation phase. This would help to confirm that the source of the  DC component is isotropic emission from the pulsar magnetosphere, as we speculated above.

\section{Discussion}

\subsection{The pulsar UVOIR spectrum}

We corrected the observed NUV and FUV magnitudes for the effects of the interstellar extinction. As was done done in Mignani et al.\ (2010a; 2012), we assumed a reddening $E(B-V)=0.2$ (see also discussion in Serafimovich et al.\ 2004). We assumed the interstellar extinction law of Fitzpatrick (1999) that gives extinction correction terms $A_{\rm NUV} = 1.62$ and $A_{\rm FUV} = 1.56$ at $\lambda=2359$\AA\ and $\lambda=1596$\AA,  respectively.  We remark that our choice of the interstellar extinction correction was made so as to not bias the comparison with the optical and near-IR fluxes of Mignani et al.\ (2010a; 2012), which were also corrected using the extinction law of Fitzpatrick (1999). By correcting the observed magnitudes according to the computed  $A_{\rm NUV}$ and $A_{\rm FUV}$, we obtain extinction-corrected fluxes of $F_{\rm NUV} = (7.88\pm0.14) \mu$Jy and $F_{\rm FUV} = (2.40\pm0.23) \mu$Jy for the NUV and FUV bands, respectively. 

\begin{figure*}
\centering
{\includegraphics[width=12cm]{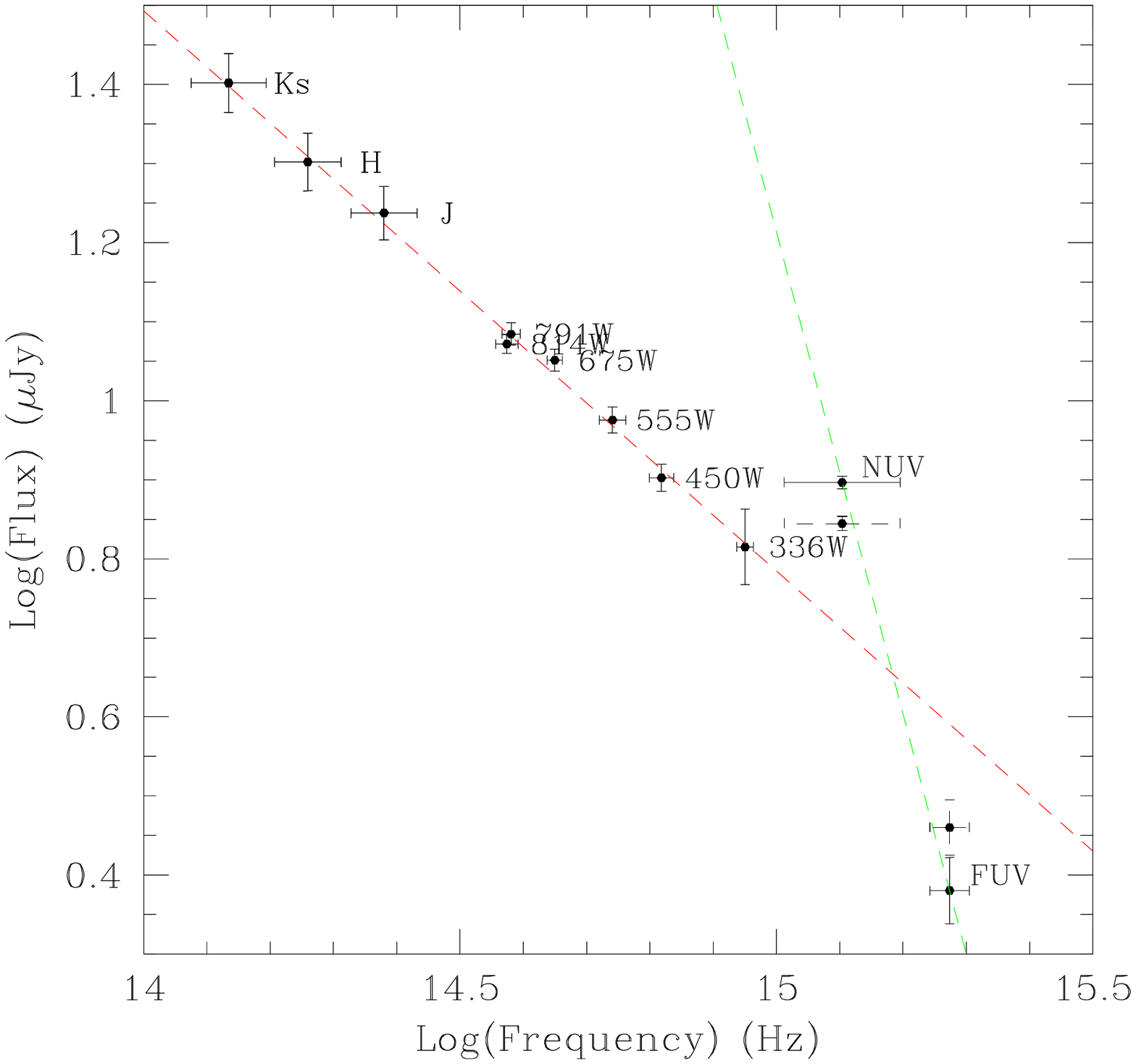}}
\caption{\label{spec} Multi-band spectrum of \psr. The {\em HST} optical flux measurements, labelled with the broad band (W) filter numbers, are taken from Mignani et al.\ (2010a) and the VLT near-IR ones (J, H, K$_{\rm s}$) from Mignani et al.\ (2012). The NUV and FUV fluxes are from the present work.  All fluxes have been obtained through imaging photometry and are integrated over the pulse-phase. The red line corresponds to the best-fit optical/near-IR PL spectrum computed by Mignani et al.\ (2012), whereas the green line indicates the PL connecting the NUV and FUV fluxes. Correction for the interstellar reddening has been applied as described in \S\ 4.1, using the extinction law of Fitzpatrick (1999).  For comparison, we also plotted the NUV and FUV fluxes (dashed lines) corrected using the extinction law for the 30 Doradus region (Gordon et al.\ 2003).   Applying this law to the optical and near-IR fluxes would not result in an appreciable difference.
}
\end{figure*}

We used these flux values to characterise the \psr\ spectrum from the near-IR to the near-UV. 
 Here we are aware that we are comparing flux measurements taken at different epochs, in particular before and after the large $\dot{\nu}$ change (Marshall et al.\ 2015). Since the effects of this event on the pulsar UVOIR flux and spectrum are unknown, such comparison must be taken with due care.
Fig.\ref{spec} shows the UVOIR spectrum of \psr\ together with the best-fit PL 
to the optical/near-IR fluxes (Mignani et al.\ 2012).  In all cases, the plotted fluxes have been computed through aperture photometry on the time-integrated images. Therefore,  the fluxes are integrated over the pulse phase, which means that they account for both pulsed and unpulsed emission components. 
As it can be seen, the NUV and FUV  fluxes are clearly incompatible with the slope of the optical/near-IR PL  ($\alpha_{\rm O,nIR}= 0.70 \pm 0.04$), with the former above and the latter below 
its extrapolation by $\sim25\sigma$ and $\sim5.5\sigma$, respectively,  suggesting a drastic turnover in the spectrum at wavelengths below 2400 \AA. In particular, the NUV and FUV fluxes are not consistent with a flattening of the optical/near-IR PL,
as one would expect from the comparison between the optical and X-ray spectra (see Fig.\ 3 of Mignani et al.\ 2010a), but instead show a steeper  PL in the UV, with spectral index $\alpha_{\rm UV}= 3.05\pm0.25$. Such a steep PL slope has never been seen in the UVOIR spectra of any other pulsar, where the spectral index is usually $\approx 0$--1 (Mignani et al.\ 2011).
Since this result is unexpected, we double-checked for possible bugs in our end--to--end procedure as follows.

Firstly, we checked that our photometry is neither affected by systematics, such as the aperture correction, nor by calibration issues, such as the zero-point definition, for which we straightforwardly applied values reported in the instrument handbook and in the image headers. As a safe measure, we verified that the tabulated aperture correction factors are consistent with those measured directly on the image and that the values of the {\sc PHOTFLAM} keywords in the image headers  were consistent with those reported in other sources (e.g., Proffitt 2006).  Therefore, we are confident that the observed magnitudes have been computed and calibrated correctly. We note that most of the fluxes plotted in Fig. \ref{spec}  have been measured with the {\em HST} and calibrated in a similar fashion, which minimises the risk of cross-calibration problems. 

As a test, we computed the pulsar photometry using methods different from those described in \S 3.1, e.g.  by employing different software tools for the photometry, by using different apertures and background areas and, consequently, different values for the aperture correction,  and obtained fully compatible magnitude values, which confirms that our results are robust and method-independent.   For consistency, we also compared our measured pulsar CRs in the NUV and FUV images with those predicted by the STIS Exposure Time Calculator (ETC). We assumed the de-reddened NUV and FUV fluxes at the peak wavelength of the F25QTZ filter, the PL connecting  these two values (Fig. \ref{spec}) as a template spectrum, an $E(B-V)=0.2$, and the zodiacal light at the \psr\ coordinates. The ETC predicts CRs of 
1.48 (NUV) and 0.053 (FUV), after accounting for the 
sky background from the PWN, 
which are fully consistent with ours. 

In principle, our photometry might have been affected by issues other than those just discussed, such as a glitch in the instrument performances.  However, no variations in the detector throughput or other anomalies have been reported in the STIS instrument science reports\footnote{{\tt  http://www.stsci.edu/hst/stis/documents/isrs}} for the time frame around our observations.    To rule out that the large difference between the pulsar NUV and FUV CRs is due to instrument effects of some sort we computed the difference between the co-aligned NUV and FUV images and found that the CR residuals for all stars in the field of view are randomly distributed above and below those for the pulsar, as one would expect if the data are free of instrument systematics. 

Finally, we have carefully checked  the applied UV extinction correction against those  derived in more recent works
and ruled out that the de-reddened NUV and FUV flux values that we derived are substantially mis-estimated. 
For instance, assuming the extinction law of Gordon et al.\ (2003), which is derived both for the LMC as a whole and, more specifically, also for the 30 Doradus region, would only marginally change the extinction correction  
in the NUV and FUV, 
whereas the extinction correction in the near-IR and the optical bands  would be undistinguishable from that derived from the extinction law of Fitzpatrick (1999).  Although this would result in a $\approx 12\%$  lower de-reddened NUV flux, and in a similarly higher de-reddened FUV flux, it would only partially account for the NUV/FUV flux difference with respect to the optical/near-IR PL  extrapolation, which would still be at  $\sim20\sigma$ and $\sim3.5\sigma$, respectively (Fig. \ref{spec}).  Since \psr\ is embedded in its supernova remnant, one can speculate of a difference in the interstellar extinction law on a more local scale, owing to a different chemical composition of the remnant with respect to the surrounding environment.  This speculation, however, cannot be easily verified with the available data, especially given the small angular extent of the remnant ($4\arcsec\times4\arcsec$).  Even so, since \psr\ lies within the 
30 Doradus nebula ($40\arcmin \times 25 \arcmin$) its effects on the interstellar extinction law
dominate over those produced by local environment fluctuations. 

Having done all the due checks, and having found no obvious bug in our procedure, we are prone to conclude that the source of the large  difference between the pulsar NUV and FUV fluxes is intrinsic to the pulsar. This suggests that, unexpected as it may be,  the abrupt turnover observed in the pulsar PL spectrum is intrinsic to the source, although the evidence  must be supported by more flux measurements in the UV.  

Thinking of a physical origin, one may speculate whether the larger (smaller)  NUV (FUV) flux with respect to the optical/near-IR PL extrapolation might be (at least partially) explained by a DC component  in the pulsar emission which is stronger in the NUV than at longer wavelengths, whereas it is almost absent in the FUV (see \S\ 3.2). If the PF were similar at all wavelengths, such a different DC  component fraction in the UV would raise the phase-integrated NUV flux above the expected value, and decrease the FUV flux, 
affecting the slope of the phase-integrated spectrum. Unfortunately, 
there are no published light curves of \psr\ obtained in bands other than V (see, Gradari et al.\ 2011 and references therein\footnote{Middleditch et al.\ (1987) indeed obtained light curves of \psr\ in the UBVRI bands but these were never published.}), so that  we do not know the value of the PF at different wavelengths. Moreover,  these light curves were all obtained with non-imaging photon counting detectors,  making it more problematic to subtract the PWN background and disentangle a genuine DC component in the pulsar emission.  Therefore, we can neither compare the pulsed nor the DC component fraction at different wavelengths. Future multi-band, high-time and high-spatial resolution observations of \psr\ with imaging photometers would be crucial to test our hypothesis. 

Another, but less likely, possibility is that of long-term variability in the pulsar UV flux  since the epochs of the optical (June--November 2007) and near-IR (October--December 2010) observations of Mignani et al.\ (2010a; 2012). Pulsars are generally known to be stable sources on long time scales and, in the case of the Crab, it has been shown that optical flux variations can be at the level of just a few milli-magnitudes per year (Sandberg \& Sollerman 2009). Larger variations, such as those observed in $\gamma$-rays for  PSR\, J2021+4026 (Allafort et al.\ 2013), cannot be ruled out a priori. However, in this case one would expect the NUV and FUV fluxes, as well as the optical and near-IR fluxes,  to vary in the same direction, unless the flux variation is accompanied by a spectrum variation. Unfortunately, there are no multi-epoch sets of UVOIR flux measurements of \psr\ to look for possible flux/spectral variations. Therefore, obtaining a new set of UVOIR flux measurements as close in time as possible to one another is the required step to search for possible long-term variability at these wavelengths. In the X-rays, no significant long-term variability had been observed  from the analysis of {\em RXTE} observations   of \psr\ (Ferdman et al.\ 2015), which cover a time span of 15.8 years. However, these observations only extend up to December 3 2011, i.e. right before the large $\dot{\nu}$ change, which occurred somewhen between December 3 and 17 2011 (Marshall  et al.\ 2015). 

As anticipated earlier in this section, we cannot rule out that an erratic phenomenon such as the large $\dot{\nu}$ change might have produced a variation in the pulsar UVOIR flux and/or spectrum, making the UV fluxes not directly comparable with the optical/near-IR fluxes, which have been measured before 2011.  Since there are neither UV flux measurements taken before the $\dot{\nu}$ change nor new optical/near-IR flux measurements taken after this event to compare with, we cannot establish whether a consequent variation in the pulsar UVOIR flux and/or spectrum ever occurred.  The same new set of UVOIR flux measurements required to look for long-term variability (see above) will also help  to determine whether the $\dot{\nu}$ change has affected the pulsar emission at these wavelengths. 
Interestingly, the last two {\em RXTE} observations of \psr, on December 17 and 31 2011, i.e. after the  large $\dot{\nu}$ change, do not show any significant difference in the X-ray flux with respect to the historical trend (Marshall et al.\ 2015). However, no other information on the X-ray flux evolution after the event has ever been reported. Therefore, we looked for X-ray variability on a more recent time frame. As a first order test, we compared the pulsar X-ray flux measured at two epochs closest in time to our near-IR (October--December 2010)  and UV (February 2017) observations using data available in  X-ray observatory archives.   To this aim, the only suitable X-ray data are those in the {\em Swift}/XRT archive, taken on November 10 2010 (13.5 ks) and February 14 2017 (1.1 ks) in PHOTON and WT mode, respectively. However, from the measured X-ray flux we found no evidence of significant variability between the two epochs. This suggests that the large $\dot{\nu}$ change did not affect the X-ray flux, as implied by the post-event {\em RXTE} observations (Marshall et al.\ 2015). A systematic analysis of all the {\em Swift}/XRT observations of \psr\ from February 17 2015 on (Marshall et al., in preparation) will allow us to look for possible X-ray flux variability over the epoch range of interest in more detail. 

The possibility that the NUV flux excess with respect to the optical/near-IR PL extrapolation is due to an emission feature centred at $\sim$ 2350 \AA, perhaps associated with an ion cyclotron line produced in the pulsar magnetosphere is, at present, no more than a speculation. The possible presence of emission/absorption features in pulsar optical spectra had been claimed for the Crab (Nasuti et al.\ 1996), Geminga (Mignani et al.\ 1998) and PSR\, B0656+14 (Durant et al.\ 2011) but the existence of these features has either not been confirmed by independent observations or it is still to be proved.  An unsubtracted spectral feature in the \psr\ supernova remnant (SNR) emitted from a region very close to the pulsar may be another
possibility. High spatial-resolution near-UV spectroscopy observations of the pulsar and of its SNR are needed to verify these two possibilities. So far, the only optical/near-UV spectrum of \psr\ was obtained by Hill et al.\ (1997)  with the Faint Object Spectrograph aboard {\em HST} but the spectral coverage (2500--5000 \AA)  did not extend to the wavelength range of interest.

\subsection{The pulsar multi-wavelength spectrum}

Regardless of the unusual PL slope in the UV, it is clear that the NUV and FUV fluxes measured for \psr\  would be incompatible with a  $\propto \nu^2$ RJ spectrum. This speaks in favour of a non-thermal (synchrotron) origin of the UV emission, as it is believed to be for the optical and near-IR  emission, powered by the pulsar rotational energy.  Under the hypothesis that the pulsar $\dot{\nu}$ change did not introduce a flux/spectrum variation (see discussion in \S 4.1), the difference in the PL slope from the optical/near-IR to the UV would, then, imply a break in the pulsar non-thermal UVOIR spectrum.

Breaks in the pulsar non-thermal UVOIR spectra are not unheard of.  Indeed, a  spectral break is observed in the Crab pulsar in the transition from the optical/near-IR to the near-UV, where the PL spectral index features a turnover from $-0.31\pm0.02$ to $0.11$ (Sollerman 2003). This break, however, is clearly not as dramatic as that observed in \psr.  The Vela pulsar, on the other hand,  features a single PL that fits the spectrum all the way from the near-IR to the near-UV (Zyuzin et al.\ 2013). Whether the presence or absence of breaks in the pulsar UVOIR spectra  depends on the characteristic age, with Vela being a factor of 10 older than the Crab, or on other pulsar parameters is not clear yet (see, e.g. Mignani et al.\ 2016b.) In the case of PSR\, B0656+14 and Geminga, the other two pulsars that had been detected in the UVOIR, the spectral break between the optical and near-UV is only due to the onset of the RJ component, which dominates over the PL one in the near-UV  (e.g., Kargaltsev \& Pavlov 2007), and not to a genuine turnover in the optical PL spectral index. 

The characterisation of the overall pulsar spectral energy distribution (see Fig.\ 3 in Ackermann et al.\ 2015) is not significantly advanced by our new NUV and FUV fluxes, given their limited spectral coverage. However, they confirm that the optical/near-IR and X-ray spectra cannot be described by a single PL, as pointed out by Mignani et al.\ (2010a) and Serafimovich et al.\ (2004) based on the optical fluxes only.   Observations at shorter UV wavelengths would help to bridge the pulsar emission in these two spectral regions. Unfortunately, {\em HST} observations cannot push the wavelength limit any further than $\approx 1100$\AA,  whereas \psr\ would have not been spatially resolved by the imaging detectors aboard the {\em Extreme Ultraviolet Explorer} (Bowyer \& Malina 1991) and has not been observed by the {\em Far Ultraviolet Spectroscopic Explorer} (Moos et al. \ 2000). A more robust characterisation of the pulsar spectrum in the near-UV  through {\em HST} spectroscopy, though, would help to make its extrapolation towards higher frequencies more accurate.

\subsection{The pulsar UV luminosity}
 
 \begin{table*}[tbh]
\begin{center}
\footnotesize{
\begin{tabular}{lrrcccccl}
\hline \noalign {\smallskip}
Name & $P_{\rm s}$ & $\tau_{\rm C}$  & $B_{\rm s}$  & $\dot{E}$ & $L_{\rm UV}$  & d  &  $\lambda_{\rm min}$--$\lambda_{\rm max}$ & Instrument \\
           & (ms) &  (kyrs) & ($10^{12}$ G) & ($10^{38}$ erg s$^{-1}$) & ($10^{34}$ erg s$^{-1}$)  & (kpc) &  (\AA) & \\ \hline
Crab$^1$		& 33.39 	& 1.26	& 3.79	& 4.50			&   1.04	& 2 & 1600--3200  & STIS/NUV-MAMA G230L \\
B0540$-$69$^2$	& 50.05 	& 1.67	& 4.98	& 4.98			&   1.27	& 48.97	& 1500--3500  & STIS/NUV-MAMA F25QTZ \\
Vela$^3$	& 89.32	& 11.3	& 3.38	& 0.069			&   1.1$\times10^{-4}$	& 0.287  &1800--3000  & STIS/NUV-MAMA F25SRF2\\
B0656+14$^4$		& 384.89	& 111	& 4.66	& 3.8$\times10^{-4}$	 &  4.2$\times10^{-5}$ & 0.288 & 1150--1700  & STIS/FUV-MAMA G140L\\
Geminga$^5$		& 237.09	& 342	& 1.63	& 3.2$\times10^{-4}$	  & 1.1$\times10^{-5}$  & 0.200 & 1800--3000  & STIS/NUV-MAMA F25SRF2\\
B1055$-$52$^6$	& 197.10	& 535	& 1.09	& 3.0$\times10^{-4}$	  & 2.3$\times10^{-5}$&  0.35 & 1350--2000  & ACS/SBC F140LP\\
B1929+10$^7$		& 226.51	& 3.1$\times10^3$	& 0.51	& 3.9$\times10^{-5}$	  & 4.6$\times10^{-6}$ & 0.33 & 1500--3500 & STIS/NUV-MAMA F25QTZ\\
B0950+08$^8$ & 253.06	& 1.75$\times10^4$	& 0.24	& 5.6$\times10^{-6}$	  &  4.3$\times10^{-6}$ & 0.262 & 1250--2000 & ACS/SBC F125LP\\
J2124$-$3358$^9$	& 4.93		& 3.8$\times10^{6}$ & 3.2$\times10^{-4}$ & 6.8$\times10^{-5}$ & 5.8$\times10^{-6}$ & 0.410 & 1250--2000 & ACS/SBC F125LP \\ 
J0437$-$4715$^{10}$  & 5.96              & 4.9$\times10^{6}$ & 5.8$\times10^{-4}$ & 3.8$\times10^{-5}$ &  4.7$\times10^{-7}$  & 0.139 & 1150--1700 & STIS/FUV-MAMA G140L \\
\hline %
\end{tabular}
}
$^1$Sollermann et al.\ (2000); $^2$this work; $^3$Romani et al.\ (2005); $^4$Shibanov et al.\ (2005); $^5$Kargaltsev et al.\ (2005); $^6$Mignani et al.\ (2010b); $^7$Mignani et al.\ (2002); $^8$Pavlov et al.\ (2017); $^9$Rangelov et al.\ (2017); $^{10}$Kargaltsev et al.\ (2004)
\caption{Pulsars detected in the UV by the {\em HST} along with the values of spin period $P_{\rm s}$, characteristic age $\tau_{\rm C}$, surface magnetic field $B_{\rm s}$, rotational energy loss $\dot{E}$, as listed in the Australia Telescope National Facility (ATNF) pulsar  catalogue  (Manchester et al.\ 2005), and the UV luminosity $L_{\rm UV}$ computed from the observed flux for  the assumed distance (d) and wavelength range ($\lambda_{\rm min}$--$\lambda_{\rm max}$). Last column reports the corresponding instrument/detector combination, either the STIS/MAMAs or the Advanced Camera for Surveys (ACS) Solar Blind Channel (SBC), and the names of the used imaging filters (F) or spectroscopy gratings (G).  For the binary millisecond pulsar PSR\, J0437$-$4715 the UV luminosity value refers to the pulsar only. }
\end{center}
\end{table*}

 \psr\ is the tenth pulsar detected in the UV by the {\em HST} (see Table 1 for a summary). The list includes the recycled millisecond pulsar PSR\, J0437$-$4715 which is in a binary system and was spectroscopically resolved from its white dwarf (WD) companion (Kargaltsev et al.\ 2004; Durant et al.\ 2012). We note that the double pulsar system PSR\, J0737$-$3039A/B  was also detected in the UV (Durant et al.\ 2014) but it was not possible to disentangle the contribution of the two pulsars in the time-integrated {\em HST} images. For this reason, neither of the two pulsars is  included in Table\, 1.
  
We computed the \psr\ UV luminosity and compared it with that of other pulsars detected in the UV.
For  \psr, the isotropic luminosity in the NUV F25QTZ filter (Riley et al.\ 2017) is $L_{\rm NUV} = 1.27 \times10^{34}$d$_{\rm LMC}^2$ erg s$^{-1}$, where $d_{\rm LMC}$ is the LMC distance in units of 48.97 kpc (Storm et al.\ 2011).  This corresponds to a fraction of $\sim 8.5 \times 10^{-5}$ of its rotational energy loss $\dot{E}$ ($1.5 \times 10^{38}$ erg s$^{-1}$). In the FUV F25QTZ filter, the luminosity is  $L_{\rm FUV} = 1.9 \times10^{33}$d$_{\rm LMC}^2$ erg s$^{-1}$ and the $\dot{E}$ fraction is correspondingly lower by a factor $\approx 6.6$.  For comparison, in the near-IR (K band) this fraction is $\sim 1.8 \times 10^{-5}$d$_{\rm LMC}^2$ (Mignani et al.\ 2012), whereas in the optical (V band) is $\sim 1.7 \times 10^{-5}$d$_{\rm LMC}^2$ (Mignani et al.\ 2010a), which means that \psr\ radiates a factor of five more energy in the NUV than at longer wavelengths.

For the other two young pulsars, Crab and Vela, the UV emission is also non-thermal. 
For the Crab, integrating its STIS/NUV-MAMA spectrum (Sollerman et al.\ 2000) over the NUV F25QTZ wavelength range gives $L_{\rm NUV} = 1.04 \times10^{34}$ erg s$^{-1}$ for a distance of 2 kpc (Manchester et al.\ 2005),  about the same luminosity as \psr.
In the case of the Crab, however, owing to its three times larger $\dot{E}$ with respect to \psr, this corresponds to an $\dot{E}$ fraction of only $\sim 2.28 \times 10^{-5}$. 
For Vela, the UV luminosity, obtained from STIS/NUV-MAMA images but in the F25SRF2 filter (Romani et al.\ 2005), which is similar to the F25QTZ one, is $L_{\rm NUV} = 1.1 \times 10^{29}$ erg s$^{-1}$ for the radio parallactic distance of 287 pc (Dodson et al.\ 2003). 
This corresponds to a fraction as low as $\sim 1.55 \times 10^{-8}$ of its $\dot{E}$ ($6.9 \times10^{36}$ erg s$^{-1}$), a factor of ten higher than the corresponding $\dot{E}$ fraction emitted in the optical, though ($\sim 1.95 \times 10^{-9}$; Moran et al.\ 2014). This shows that also in the UV, like in the optical, Vela
emits a lower fraction of its $\dot{E}$ with respect to the very young pulsars Crab and \psr.  

Strictly speaking, a direct comparison with the UV luminosity of the middle-aged pulsars PSR\, B0656+14, 
Geminga, 
PSR\, B1055$-$52 
(Shibanov et al.\ 2005; Kargaltsev et al.\ 2005; Mignani et al.\ 2010b) would not be very informative because of the difference in the underlying emission mechanisms. For the middle-aged pulsars the UV emission is dominated by thermal radiation from the cooling neutron star surface and not by non-thermal radiation from the neutron star magnetosphere, as in the case of the young Crab, Vela, and \psr. The UV emission is also thermal  for the $\sim 17.5$ Myr-old PSR\, B0950+08 (Pavlov et al.\ 2017) and  the $\sim 4.9$ Gyr-old recycled millisecond pulsar PSR\, J0437$-$4715 (Kargaltsev et al.\ 2004),
whereas for both 
the $\sim 3$ Myr-old PSR\, B1929+10 (Mignani et al.\ 2002) 
and the $\sim 3.8$ Gyr-old recycled millisecond pulsar PSR\, J2124$-$3358 (Rangelov et al.\ 2017)
 the available spectral information is not sufficient to determine whether  the UV emission is thermal or non-thermal. 
 As a further complication,  in many cases the UV flux values reported in the literature have been obtained with different {\em HST} instruments, different techniques (imaging photometry or spectroscopy) and in different wavelength ranges, which makes the inferred UV  luminosities 
 not directly comparable to each other.

\begin{figure*}
\centering
{\includegraphics[width=10cm, angle=270]{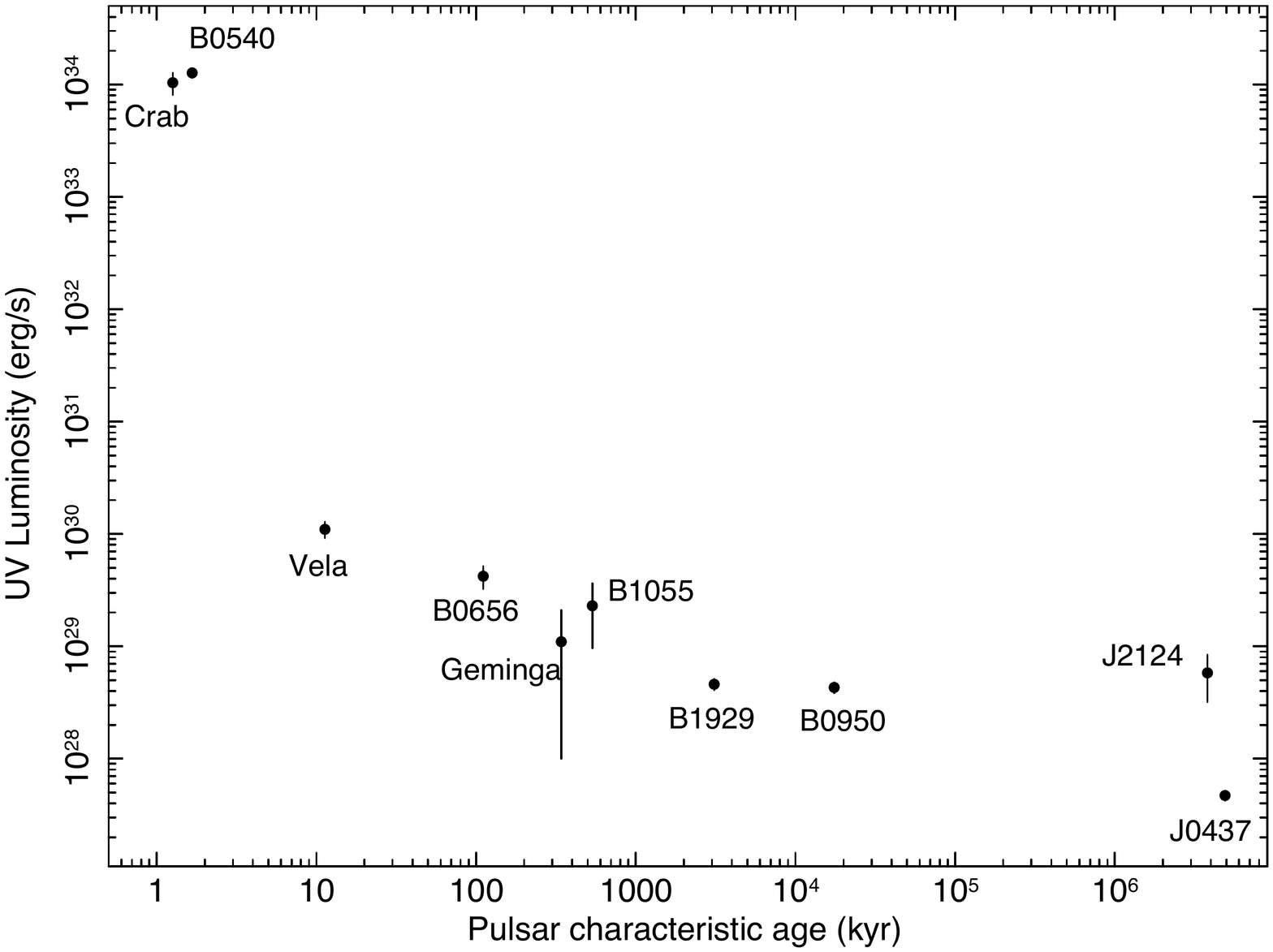}}
\caption{\label{luvev} UV luminosity plotted as a function of characteristic age for all pulsars listed in Table 1. Luminosity uncertainties account for both distance and photometry errors.
}
\end{figure*}

Therefore, given the very small sample (Crab, Vela, and \psr) it is difficult to speculate about possible trends in the pulsar non-thermal UV luminosity as a function of the pulsar parameters, e.g. the surface magnetic field $B_{\rm S}$ or the rotational energy loss $\dot{E}$. On the other hand, for the other pulsars the thermal UV luminosity is expected to be insensitive to these parameters, if emitted from a large fraction of the neutron star surface and not from hot polar caps, but to be sensitive to the temperature of the emitting region.  Since in the UV we see only the RJ part of the thermal spectrum, the brightness temperature $T_{\rm B}$ is parametrised by the second power of the ratio between the pulsar distance and the radius of the emitting region, which cannot be easily determined. Indeed, in the lack of modulations at the pulsar spin period in the thermal UV emission the only hard limit is imposed by the neutron-star radius predicted by different equations of state.  This means that deriving a temperature value for comparison with, e.g. neutron star cooling models comes with significant uncertainties.
For the sake of completeness, in Table 1 we reported the luminosity values for all pulsars detected in the UV regardless of the nature of the emission.
For illustrative purposes, Fig. \ref{luvev} shows the pulsar UV luminosity $L_{\rm UV}$ as a function of the characteristic age $\tau_{\rm C}$ for all pulsars in Table 1.  As it can be seen, the UV luminosity quickly drops for ages above $\sim$ 10 kyrs, i.e. about that of the Vela pulsar, and the trend more or less flattens above $\sim$ 100 kyrs.  This is expected since the contribution of the UV non-thermal emission becomes less important  for pulsars older than $\sim$ 100 kyrs.
A similar trend has been found for the pulsar optical luminosity (e.g., Zharikov \& Mignani 2013), marking also in this case the difference between young and middle-aged/old pulsars, although for the latter the contribution of the non-thermal emission can still be important.
 
\subsection{The pulsar UV and optical light curves}

\begin{figure*}
\centering
{\includegraphics[width=16cm]{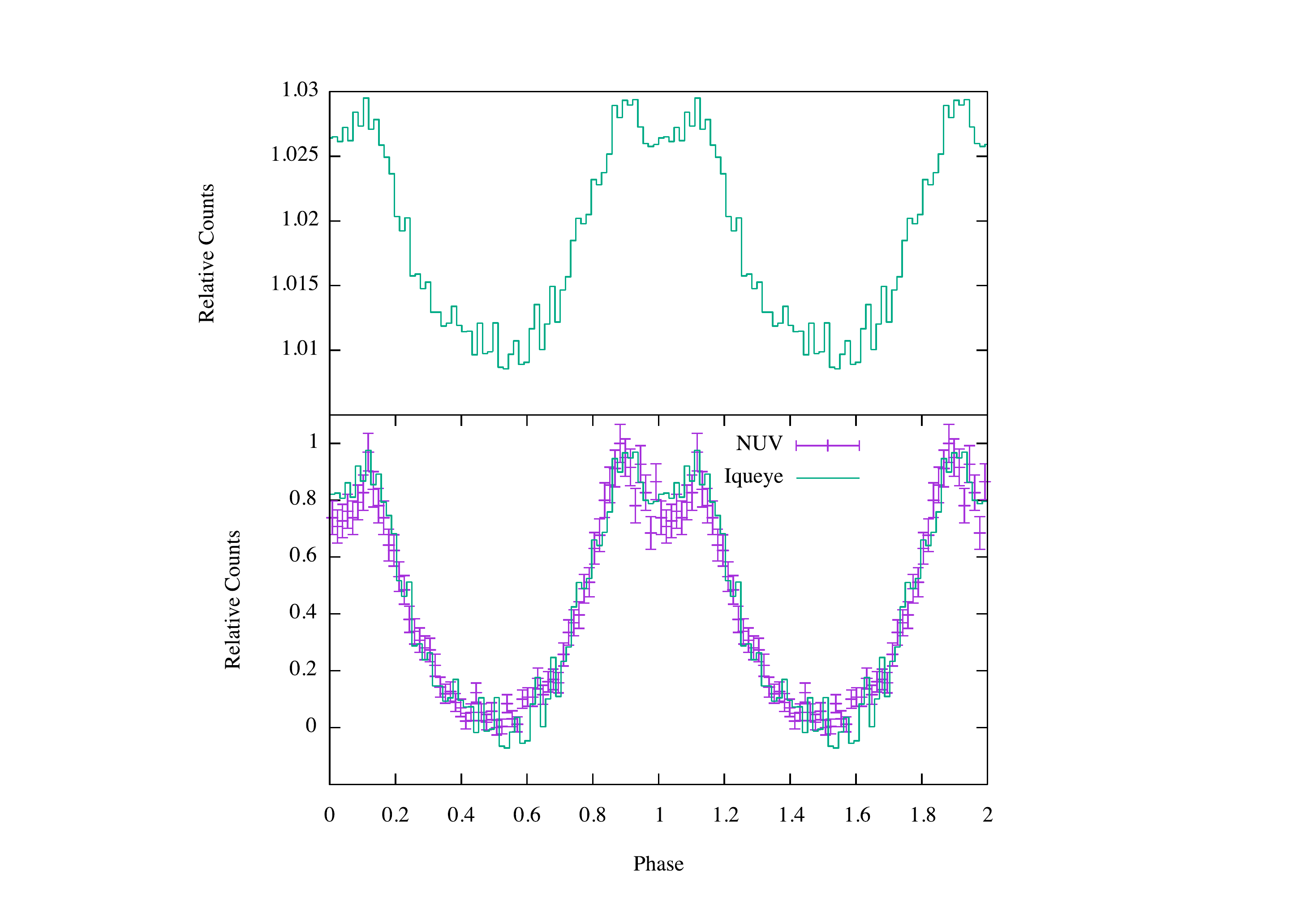}}
\caption{\label{ouv} (top) Optical light curve taken through the white band filter reconstructed from a re-analysis of the Iqueye data of Gradari et al.\ (2011).  Two cycles are shown for clarity. Since these data were obtained back in 2009, i.e. before the large change in the pulsar $\dot{\nu}$ (Marshall et al.\ 2015),  the light curve has been folded using the ephemeris derived from {\em RXTE}/PCA observations over the period May 16 2008--December 3 2011 (Ackermann et al.\ 2015). The  third peak tentatively seen in the Iqueye light curve published in Gradari et al.\ (2011) is not visible here; see \S\ 4.4 for details.  (bottom) NUV light curve (magenta line) from Fig. \ref{lc} superimposed on that constructed from the Iqueye data set (green line). 
}
\end{figure*}

Being close in wavelength, it is natural to compare first the UV light curves of \psr\ to those in the optical band.\footnote{\psr\ has been detected in the near-IR (Mignani et al.\ 2012), but pulsations in these band have not yet been measured.} 
In the optical, its most recent light curve has been published by Gradari et al.\  (2011) based on data obtained with the Iqueye instrument  (Naletto et al.\ 2009) at the ESO New Technology Telescope (NTT) on January and December 2009. Iqueye observations taken during the same observing runs were also used to produce an updated optical light curve of the Vela pulsar (Spolon et al.\ 2019).  The \psr\ light curve profile clearly revealed a two-peak structure, with the two peaks separated in phase by $\sim 0.3$, in agreement with all the \psr\ optical light curves reported in the literature (see Gradari et al.\ 2011 and refs. therein). Fig. \ref{ouv} (top) shows the  light curve built from the Iqueye data of Gradari et al.\ (2011),
as published in  Fig.\ 2 of Ackermann et al.\ (2015).

We note that Gradari et al.\ (2011) found possible evidence (at the $\approx 3\sigma$ level) of a third peak in the light curve interposed between the two main peaks (see their Fig.\ 1) but this is not visible in Fig. \ref{ouv} (top). 
The reason behind this discrepancy, never addressed so far, is that the Iqueye data have been  fully re-processed by Ackermann et al.\ (2015) with an upgraded version of the data reduction software, which improved  the determination of the photon time of arrival.  Another, and likely more important, reason is the use of a different ephemeris for the light curve folding. Gradari et al.\ (2011) did not have simultaneous ephemeris available and then folded and aligned the data on the basis of their own period measurements,
whereas Ackermann et al.\ (2015) used the ephemeris
derived from observations with the {\em Rossi X-ray Timing Explorer} ({\em RXTE}) Proportional Counter Array (PCA), between May 16 2008 and December 3 2011 (MJD 54602--55898).
Therefore,  the third peak seen in the Iqueye data by Gradari et al.\ (2011) was probably an artifact of the data analysis.  Interestingly, this peak might correspond to  that possibly seen between the two main peaks in the NUV light curve  (Fig. \ref{lc}, top), whose significance, however, is also marginal (see \S\ 3.2).  Although two coincidences may represent  a clue, only follow-up optical/UV observations, possibly with different telescope/instrumental set-ups, can provide more convincing evidence of the existence of this putative third peak.  
Confirming its existence would  unveil a more complex light curve morphology than initially thought, which might encode thus far missing information on the pulsar viewing and beaming angles and the structure of the optical/UV emission cone.  

As it can be seen from 
the comparison between Fig. \ref{lc} and Fig. \ref{ouv} (top), 
the NUV/FUV light curve profiles bear resemblance to the optical one, with two peaks separated in phase by approximately the same amount. This resemblance is more noticeable for the NUV light curve, as shown by a direct comparison in Fig. \ref{ouv} (bottom), where the two peaks have similar relative intensities, as in the optical light curve.  This means that the difference in the pulsar PL spectrum between the optical and the UV (\S\ 4.1) did not affect the light curve profile.
The UV and optical light curves  are also aligned in phase, although they correspond to different epochs (2017 and 2009) and have been folded using different sets of ephemerides owing to the large change in the pulsar $\dot{\nu}$ that occurred between December 3 and December 17 2011  (Marshall et al.\ 2015).  The UV light curves have been folded using the February 17 2015--March 28 2018 {\em Swift}/XRT ephemeris (Marshall et al., in preparation),  as explained in \S\ 3.2,  whereas the optical light curve has been folded using the May 16 2008--December 3 2011 {\em RXTE}/PCA ephemeris (Ackermann et al.\ 2015), as explained above. 
A question then arises about whether the observed alignment is real or whether the large $\dot{\nu}$ change of December 2011 might have introduced a systematic phase offset,  so that the 2017 (UV) and 2009 (optical) light curves of \psr\ would not be directly comparable to each other. This is a key point for our analysis, aimed at determining whether the observed break between the optical/near-IR and UV PL spectra  has consequences  not only on the profile of the optical and UV light curves but also on their alignment in phase. 
Since there are no UV (optical) light curves of \psr\ obtained before (after) December 3 2011
 for a direct comparison  we cannot directly clarify this point.  In the X- and $\gamma$-rays, however, the comparison between 
  light curves
  obtained before and after the large $\dot{\nu}$ change
 does not show any obvious misalignment 
 (see \S\ 4.5).   
This suggests that this event did not introduce a major phase offset, at least at high energies, and we can reasonably assume that this is also the case for the optical and UV, although our hypothesis can only be confirmed by new optical timing observations for comparison with those of Gradari et al.\ (2011). %

Therefore, the difference in slope of the pulsar PL spectrum between the optical/near-IR and the UV,  if  intrinsic to the UVOIR spectrum  and not ascribed  to spectral/flux variability (see discussion in \S\ 4.1), would have no consequences on the phase alignment of the optical and UV light curves.  
Their close resemblance (Fig. \ref{ouv}, bottom) independently supports the evidence based on the spectrum that, as in the optical,  the UV radiation is of magnetospheric origin. In particular, it suggests that the optical and UV radiation have a very similar emission geometry, whereas the almost perfect phase alignment between the peaks suggests that the emission region in these two bands is most likely the same.

\subsection{The pulsar multi-epoch light curves}

As anticipated in the previous section, here we describe the results of the comparison between  the X and $\gamma$-ray light curves of \psr\ obtained at different epochs, carried out in this work for the first time.

\begin{figure*}
\centering
{\includegraphics[width=10cm, angle=0]{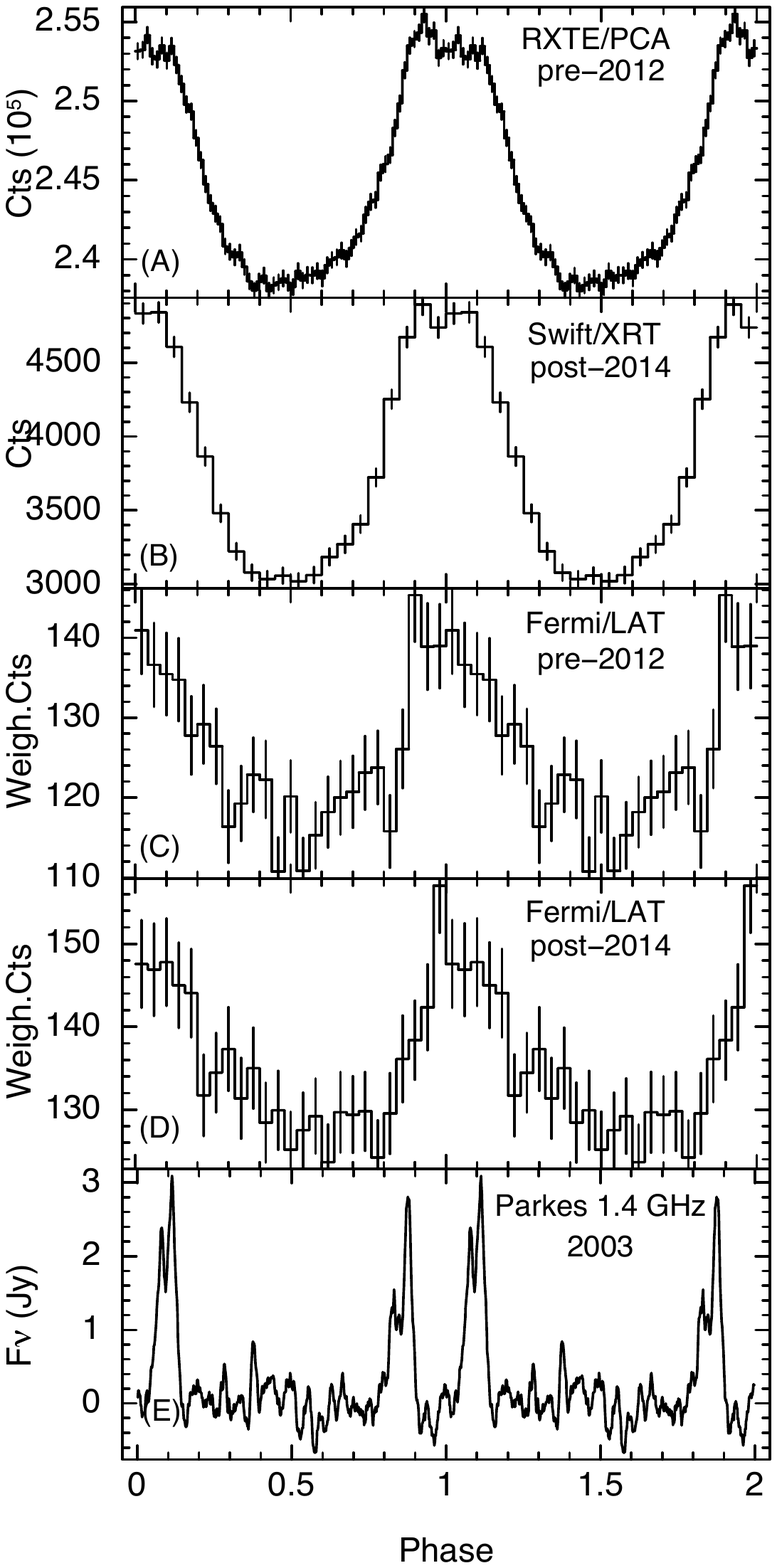}}
\caption{\label{mlc} 
X and $\gamma$-ray light curves of \psr\ obtained from data taken before  (pre-2012) and after (post-2014) the large $\dot{\nu}$ change occurred between  December 3 2011 and December 17 2011 (Marshall et al.\ 2015). From top to bottom:  {\em RXTE}/PCA X-ray light curve (MJD 54602--55898), {\em Swift}/XRT X-ray light curve (MJD 57070--58205), 
{\em Fermi}/LAT $\gamma$-ray light curves  (MJD 54689--55898 and MJD 57070--58270).  Both the pre-2012 and post-2014 LAT light curves have been built using using the photon weighting method described in Bruel et al.\ (in preparation).  In all cases, the light curves have been built by integrating all the data collected over the time intervals reported above in parentheses.  The pre-2012 {\em RXTE}/PCA  and {\em Fermi}/LAT light curves are based on the same data as used in Ackermann et al.\ (2015). In the first and third panels, the light curves have been folded using the pre-2012 ephemeris obtained from the {\em RXTE}/PCA  data set (Ackermann et al.\ 2015), whereas in the second and fourth panels they have been folded using the post-2014 ephemeris  obtained from the {\em Swift}/XRT  observations (Marshall et al., in preparation).  The bottom panel shows, as a reference, the  radio light curve at 1.4 GHz obtained from Parkes in August 2003 (Johnston et al.\ 2004), which is the same as shown in Ackermann et al.\ (2015). }
\end{figure*}

 Fig. \ref{mlc}  shows two sets of X and $\gamma$-ray light curves of \psr.  The first set (panels a and c) corresponds to an epoch range antecedent to the beginning of 2012 (hereafter "pre-2012"), i.e. {\em before} the large $\dot{\nu}$ change  that occurred between December 3 2011 and December 17 2011 (Marshall et al.\ 2015).  In particular, Fig. \ref{mlc}a shows the {\em RXTE}/PCA X-ray light curve built by integrating all data taken between May 16 2008 and December 3 2011 (MJD 54602--55898), whereas Fig. \ref{mlc}c shows the {\em Fermi}/LAT $\gamma$-ray light curve built from contemporary data taken between August 5 2008 and   December 3 2011 (MJD 54682--55898).   Both the {\em RXTE}/PCA and {\em Fermi}/LAT data are the same as used in Ackermann et al.\ (2015). In both panels, the light curves have been folded using the pre-2012 ephemeris obtained from the full (MJD 54602--55898) {\em RXTE}/PCA  data set, as done in Ackermann et al.\  (2015). 
 The second set (panels b and d) corresponds to an epoch range subsequent to the end of 2014 (hereafter "post-2014"), i.e. {\em after}  the large $\dot{\nu}$ change. The X and $\gamma$-ray light curves are built by integrating all data taken with the {\em Swift}/XRT  between February 17 2015 and March 28 2018 (MJD 57070--58205)  and with the {\em Fermi}/LAT between February 17 2015 and June 1 2018 (MJD 57070--58270), respectively.  Therefore, both data sets cover the epoch range around our {\em HST} observations. The {\em Swift}/XRT  data are the same as described in \S\ 3.2 and have been partially published in Marshall et al.\ (2015, 2016), whereas the new \textit{Fermi}/LAT data have not been published before.  
In both panels,  the light curves have been folded using the post-2014 ephemeris  obtained from the full (MJD 57070--58205)  {\em Swift}/XRT data set (Marshall et al., in preparation), which has been used to fold our {\em HST}/STIS light curves (\S\ 3.2).

For consistency, we analyzed both the pre-2012 and post-2014  \textit{Fermi}/LAT data sets, which cover virtually identical time spans ($\approx$ 1200 d), using exactly the same procedure. In particular,  we produced $\gamma$-ray light curve profiles for PSR\, B0540$-$69 by using Pass 8 Source class events, analyzing 
photons with energies above 0.1 GeV and with reconstructed directions within 8$^\circ$ of the pulsar. Events with zenith angles above 105$^\circ$ were rejected, to limit the contamination caused by the Earth's limb. 
Phase calculations were carried out using the \textit{Fermi} plugin (Ray et al.\ 2011) of \textsc{TEMPO2} (Hobbs et al.\ 2006). In order to improve the signal-to-noise ratio of the $\gamma$-ray light curve for both the pre-2012 and post-2014 time intervals, we assigned weights to the individual photons using the weighting method described in Bruel et al.\ (in preparation). The weights give the probabilities that the individual photons originated from PSR\, B0540$-$69. We find that using $\log_{10} E_r = 3.2$ where $E_r$ is the reference energy in MeV of the weighting algorithm (see Bruel et al.\ for a description) optimizes the signal-to-noise ratio of the profiles. 
The pre-2012  \textit{Fermi}/LAT light curve profile (Fig. \ref{mlc}c), built using the weighting method described above, is consistent with that presented in Ackermann et al.\ (2015), indicating that we did not introduce any bias or systematic effect.

As it can be seen from the comparison between the {\em RXTE}/PCA and {\em Swift}/XRT light curves (Fig. \ref{mlc}a,b), the X-ray light curve profile has not changed appreciably 
between the two epochs. The light curves are qualitatively similar, both featuring two peaks superimposed on a broad pulse, although the {\em RXTE}/PCA light curve benefits from a better statistics. Furthermore, the two light curves appear to be essentially aligned in phase. This is also true for the  pre-2012 and post-2014 {\em Fermi}/LAT light curves  (Fig. \ref{mlc}c,d).  No significant variation is observed between the pre-2012 and post-2014 LAT light curve profiles either, although the former seems to feature a more pronounced emission in the phase interval corresponding to the off-pulse region. This off-pulse emission component was already noticed by Ackermann et al.\ (2015), who could not  determine whether this was associated with the pulsar or its PWN/SNR or with  residual emission from the LMC.  A more detailed analysis of the $\gamma$-ray data, which is beyond the goals of this work, is needed to determine how the significance of this excess depends on the modelling and subtraction of the background, on the binning used in the light curve,  and, peraphs, on the count statistics.

Ours is  the first high-energy follow-up of \psr\ after its large $\dot{\nu}$ change (Marshall et al.\ 2015).
The above comparison shows that  this event did not introduce either a major phase offset or profile variation in the pulsar light curves, implying that the emission geometry did not change appreciably between the two explored epoch ranges.
Furthermore, the comparison between two pre-2012 and post-2014 {\em Swift}/XRT observations (\S\ 4.1) shows that the $\dot{\nu}$ change did not introduce a variation in the X-ray flux. 
A qualitative comparison of the counts in the pre-2012 and post-2014 {\em Fermi}/LAT light curves (Fig. \ref{mlc}c,d),  which are directly comparable to each other (see above), suggests that no variation has occurred in the $\gamma$-ray flux  either. This conclusion will be verified by an in-depth analysis of the two {\em Fermi}/LAT data sets, whose results will be published in a follow-up paper.
Therefore, based on current evidence, we conclude that the event had no consequence on the pulsar high-energy emission properties.

\subsection{The pulsar multi-wavelength light curves}

Here, we briefly describe the comparison between the optical/UV and the X/$\gamma$-ray  light curves  and discuss the implications on our understanding of the pulsar emission geometry.  

In general, the UV light  curves of \psr\  (Fig. \ref{lc}) fit very well the picture of a multi-wavelength light curve profile characterised by a broad pulse with two peaks,  as emerged from optical  (Fig. \ref{ouv}) and X-ray (Fig. \ref{mlc}a,b) observations; see Fig.\ 2 of Ackermann et al.\  (2015).  There is no noticeable shift in  the pulse phase across the UV/optical/X-ray light curves and there is no evidence of a variation  either in the peak separation or  in the relative peak intensity as a function of energy.
These two peaks are not apparent in the $\gamma$-ray light curve (Fig. \ref{mlc}c,d), though, possibly because of the lower count statistics and larger errors, whereas the alignment in phase  with the UV/optical/X-ray light curves is maintained.
For comparison, Fig. \ref{mlc}d shows the \psr\ radio light curve at 1.4 GHz obtained in August 2003 from the Parkes radio telescope  (Johnston et al.\ 2004), also shown in Ackermann et al.\ (2015). Unfortunately, owing to the pulsar faintness in radio, it was not possible to obtain more recent observations. Indeed,  the radio light curve shown in Fig. \ref{mlc}d,  the last to be published, was built exploiting the occurrence of
18 bright giant radio pulses (Johnston et al.\ 2004).   As it can be seen, the radio  light curve profile,  with two well-distinct narrow and structured peaks,  is visually different from those at higher energies, suggesting a different emission geometry.  Interestingly, the radio peaks are essentially aligned in phase  with those observed in the X-ray, optical and UV light curves, assuming in this last case no major phase offset in radio after the large $\dot{\nu}$ change (Marshall et al.\ 2015). 

Such a self-similar and phase-aligned light curve profile across different energy ranges is quite remarkable if compared to other young pulsars, such as Vela (Romani et al.\ 2005).  In particular, 
while the similarity of the light curve profiles suggests a similar emission geometry, their 
alignment in phase suggests that the pulsed multi-wavelength emission in \psr\  originates from regions very close to one another in the neutron star magnetosphere.  A  more or less self-similar and phase-aligned light curve profile across the optical--to--$\gamma$-ray energy ranges is also observed in the Crab pulsar.  These are the only two pulsars featuring this distinctive characteristic,  which strengthens the link between \psr\ and its "twin". However, the still-limited number of pulsars seen to pulsate from the optical to the $\gamma$-rays (five; \S\ 3.2) makes it difficult to establish whether such an alignment is the rule or the exception.  
Detecting multi-wavelength pulsations from a larger pulsar sample is obviously needed to address this issue.
The middle-aged ($\tau_{\rm C} \sim 0.5$ Myrs) pulsar PSR\, B1055$-$52  ($P_{\rm s} \sim 197$  ms), detected in radio, optical, UV, X-rays and $\gamma$-rays but not yet in the near-IR (Mignani et al.\ 2010b),  is the most obvious target to search for UV pulsations\footnote{The pulsar is at $\sim$4\arcsec\ from a 14.6 magnitude star and it has been detected in the optical with the {\em HST} which, however, has no instrument for high-time resolution observations above 3000\AA\ after the decommissioning of the High Speed Photometer. } and compare the light curve profile with those already measured in radio, X-rays and $\gamma$-rays.
The other young ($\tau_{\rm C} \sim 4900$ yrs) LMC pulsar PSR\, J0537$-$6910 ($P_{\rm s} \sim 16$  ms) in the N157B SNR would be, ideally, the best target owing to an $\dot {E}  \sim 4.9 \times 10^{38}$ erg~s$^{-1}$, the largest in the pulsar family. However, so far it has eluded detections at energies other than in X-rays, where it was discovered as an X-ray pulsar (Marshall et al.\ 1998), and in $\gamma$-rays, although  pulsations have not yet been detected in the latter case (Ackermann et al.\ 2015),  whereas no radio, optical, UV  counterpart has been found despite multiple attempts   (e.g., Crawford et  al.\  2005; Mignani et al.\ 2005; Mignani et al.\ 2007). 

Searching for near-IR pulsations from \psr, never detected in any pulsar other than the Crab (Eikenberry et al.\ 1997), would be the next goal towards completing the  multi-wavelength picture for this source and allow for a full band--to--band comparison with its "twin".  In particular, high-time and spatial resolution near-IR observations would  help to disentangle the light curve contribution from a possible DC component in the pulsar emission from that of the PWN background.  From the comparison with our UV light curves, and with those in the optical band, also to be obtained through high-time and spatial resolution observations, it will then be possible to determine whether the DC component fraction remains constant or evolves with wavelength.    As discussed in \S\ 4.1, a different DC component contribution in the UV with respect to the optical/near-IR could help to explain the abrupt UV turnover in the pulsar spectrum.

\section{Summary and Conclusions}

Using the STIS-MAMAs aboard {\em HST}, we detected the LMC pulsar \psr\ in two near-UV bands centred around 2350\AA\ (NUV) and 1590\AA\  (FUV) and measured pulsations at the pulsar spin period in both bands. This is the first time that \psr\ has been detected and seen to pulsate in the UV. 
Aside from the radio,  \psr\ is  now one of the five pulsars  (counting the radio-quiet Geminga)  detected in five different energy bands (near-IR, optical, UV, X-rays, $\gamma$-rays)  and seen to pulsate in at least four of them. 
PSR\, B1055$-$52, detected in all these bands but the near-IR (Mignani et al.\ 2010b),  could be next in the list.
We also detected the \psr\  PWN  in our NUV observation, with a morphology similar to that observed in the optical and near-IR, but not in the FUV, which indicates a sharp decrease of the PWN surface brightness at shorter wavelengths.

 The UV light curves of \psr\ feature a prominent broad pulse with two peaks very close in phase,  similarly to that observed in the  optical and X-rays.     A significant DC component is also observed in the NUV light curve, possibly associated with unpulsed isotropic emission from the neutron star magnetosphere.
 Like in the Crab pulsar, the UV light curves are also aligned in phase with those in the radio, optical, X and $\gamma$-rays, although these are not always contemporary to one another.
Thus,  it seems that the large change in the spin frequency derivative $\dot{\nu}$  that occurred at the end of 2011 (Marshall et al.\ 2015) did not introduce a major phase offset, as we demonstrated, at least at high energies, from  the qualitative comparison between the {\em RXTE}/PCA, {\em Swift}/XRT and {\em Fermi}/LAT light curves of \psr\ obtained before and after the event.
 The pulsar UV fluxes clearly deviate from the extrapolation at shorter wavelengths of the best-fit PL to the optical/near-IR fluxes ($\alpha_{\rm O,nIR} \sim 0.7$; Mignani et al.\ 2012). Under the hypothesis of no long-term flux variability, this would point at an abrupt steepening of the PL spectrum in the UV ($\alpha_{\rm UV} \sim 3$).  This has not yet been  observed in other pulsars  and its explanation remains a challenge.  

 More {\em HST} observations are necessary to independently confirm the difference in the pulsar PL slope in the UV and obtain a better characterisation of the pulsar spectrum at wavelengths below 3000\AA, which so far is based on our two flux measurements only.  This would require multi-band UV photometry with the Advanced Camera for Survey or, better yet, UV spectroscopy observations with either the STIS or the Cosmic Origin Spectrograph.  In particular, time-resolved UV spectroscopy with the STIS  (125 $\mu$s resolution) would enable one to better decouple the  spectrum  of the pulsar from that of its PWN and look, for the first time, for possible variations in the pulsar PL spectrum as a function of the neutron star rotation phase. This would be important to track possible differences in the properties of the emitting particles (density, velocity) in different regions of the neutron star magnetosphere, which are seen as the neutron star rotates.
 
\acknowledgments
We thank the anonymous referee for his/her careful review of our manuscript.  RPM acknowledges financial support from an INAF "Occhialini Fellowship" and thanks Ralf Siebenmorgen (ESO) for advice on the interstellar extinction correction. We thank our {\em HST} Program Coordinator, Denise Taylor (STScI), for constant support in the observation planning and Tony Sohn (STScI) for checking the STIS data status. We thank Patrizia A. Caraveo for comments and suggestions. B.R. acknowledges financial support from the National Science Center Grant DEC-2011/02/A/ST9/00256. Based in part on observations made with the ESO NTT telescope at the La Silla Paranal Observatory under programmes 082.D-0382 and 084.D-0328(A).  The \textit{Fermi} LAT Collaboration acknowledges generous ongoing support from a number of agencies and institutes that have supported both the development and the operation of the LAT as well as scientific data analysis. These include the National Aeronautics and Space Administration and the Department of Energy in the United States, the Commissariat \`a l'Energie Atomique and the Centre National de la Recherche Scientifique / Institut National de Physique Nucl\'eaire et de Physique des Particules in France, the Agenzia Spaziale Italiana and the Istituto Nazionale di Fisica Nucleare in Italy, the Ministry of Education, Culture, Sports, Science and Technology (MEXT), High Energy Accelerator Research Organization (KEK) and Japan Aerospace Exploration Agency (JAXA) in Japan, and the K.~A.~Wallenberg Foundation, the Swedish Research Council and the Swedish National Space Board in Sweden.
Additional support for science analysis during the operations phase is gratefully acknowledged from the Istituto Nazionale di Astrofisica in Italy and the Centre National d'\'Etudes Spatiales in France. This work performed in part under DOE Contract DE-AC02-76SF00515.

{\it Facilities:} \facility{{\em Hubble Space Telescope}, {\em Neil Gehrels Swift Observatory}, {\em Fermi Gamma-ray Space Telescope}, Iqueye}

\end{document}